\documentclass[aps,prd,reprint,a4paper,showpacs,nofootinbib,superscriptaddress]{revtex4-2}

\usepackage{bbm}
\usepackage{amsmath}
\usepackage{graphicx}
\usepackage{float}
\usepackage{amssymb}
\usepackage{subfigure}
\usepackage{amssymb}
\usepackage{hyperref}
\usepackage{epstopdf}

\usepackage[utf8]{inputenc}
\hypersetup{
    colorlinks=true,
    linkcolor=red,
    citecolor=blue,
}
\usepackage{color}
\usepackage[T1]{fontenc}
\usepackage{txfonts}
\usepackage{mathrsfs}
\usepackage{latexsym}
\usepackage{epsfig}
\usepackage{epstopdf}
\usepackage{epstopdf}
\usepackage{graphicx}
\usepackage{amssymb}
\usepackage{amsmath}
\usepackage{dcolumn}
\usepackage{bm}
\usepackage{color}
\usepackage{comment}
\usepackage{xcolor}
\usepackage{dsfont}
\usepackage{orcidlink}

\begin{document}
\title{\bf Unified Mass-Scaled QPO Signatures of Kerr–Sen Black Holes from Stellar-Mass to Supermassive Sources}

\author{Orhan~Donmez\orcidlink{0000-0001-9017-2452}}
\email{orhan.donmez@aum.edu.kw}
\affiliation{College of Engineering and Technology, American University of the Middle East, Egaila 54200, Kuwait}

\author{G. Mustafa\orcidlink{0000-0003-1409-2009}}
\email{gmustafa3828@gmail.com  }
\affiliation{Department of Physics, Zhejiang Normal University,
Jinhua 321004, China}
\affiliation{Research Center of Astrophysics and Cosmology, Khazar University, Baku, AZ1096, 41 Mehseti Street, Azerbaijan}

\begin{abstract}
In this study, we numerically investigate Bondi--Hoyle--Lyttleton (BHL) accretion around Kerr--Sen black holes and examine how the charge-related deformation of the spacetime affects the shock-cone morphology, the variation of the mass-accretion rate, and the quasi-periodic oscillation (QPO)-like temporal behavior. The relativistic BHL flow is solved numerically in the equatorial plane for two different black-hole spin parameters, $a=0.9M$ and $a=0.5M$. From the numerically computed mass-accretion-rate signal, we calculate the power spectral density (PSD) and perform multi-component Lorentzian fits in order to identify the dominant QPO-like modes excited around the black hole. The results show that the Kerr--Sen deformation shifts the characteristic frequencies, changes the coherence properties of the oscillation modes, and produces near-resonant harmonic structures close to 3:2 and 2:1. By using inverse mass scaling, the numerically computed frequencies are compared with observed QPOs from stellar-mass, intermediate-mass, and supermassive black-hole systems. In particular, a reasonable agreement between the numerical simulation results and the observational results is found for the sources GRS~1915+105, IGR~J17091--3624, M82~X--1, NGC~5408~X--1, RE~J1034+396, 1H~0707--495, and ESO~113--G010. This comparative analysis indicates that Kerr–Sen black-hole shock-cone oscillations may provide a unified framework for interpreting timing features over a broad range of black-hole masses and may additionally contribute to constraining the mass and spin parameters of sources whose properties are not yet fully established observationally. These findings further imply that combined hydrodynamical and timing diagnostics constitute a promising approach for assessing the extent to which deviations associated with the Kerr–Sen geometry can be empirically distinguished from those of the Kerr spacetime.\\
\textbf{Keywords}: Black-hole accretion; quasi-periodic oscillations; observational black-hole timing; Kerr–Sen black holes; power spectral density.
\end{abstract}

\maketitle

\date{\today}

\section{Introduction}
\label{Intro}

General relativity (GR) has been remarkably successful in describing gravity over a wide range of scales, from the Solar System to compact astrophysical systems. In GR, black holes appear as exact solutions of the Einstein field equations and constitute one of the most important laboratories for testing gravity in the strong gravitational field. The rotating black-hole solution was first obtained by Kerr in 1963 \cite{Kerr:1963ud}. This rotating black hole has since become the standard background for modeling astrophysical black holes, accretion flows, relativistic jets, and high-energy timing phenomena \cite{Chandrasekhar:1984siy}. Although GR has passed many precision tests, the near-horizon region of rotating black holes remains an important region in which possible deviations from the Kerr spacetime can be investigated. In particular, after the black-hole shadow observations by the Event Horizon Telescope, such deviations have started to attract more attention \cite{EHTM87_2019, EHTSgrA_2022}.\\

\noindent
Modified theories of gravity provide a natural framework for revealing possible deviations or corrections to GR, especially in the strong gravitational field. These theories have been studied extensively due to the emergence of several open problems. At the same time, modified gravity may also establish a connection between gravity and quantum theory \cite{Stelle:1976gc}. Modified gravity can help us better understand the nature of compact objects and may also appear as an alternative gravitational framework capable of producing solutions in cases where astrophysical black holes may carry observable signatures beyond the standard Kerr solution \cite{Berti:2015itd, CANTATA:2021asi}. String theory is known as one of the most popular candidate theories to unify all four fundamental forces in nature \cite{zwiebach2004first}. Simiwhich was formulatedlar to GR, string theory also admits the black hole solutions \cite{lust1989lectures, hartle2003gravity}. A few years after the Kerr black hole solution, Sen applied a solution-generating technique to the Kerr black-hole solution and obtained a charged rotating black-hole solution in the low-energy limit of heterotic string theory. This solution is known as the Kerr–Sen black hole \cite{houri2010generalized, gwak2017cosmic, xavier2020shadows, Roy:2025hdw}. The Kerr–Sen black hole solution differs from the Kerr black hole solution in that it is characterized by three parameters which are mass, angular momentum, and charge. Moreover, the Kerr–Sen solution arises in the low-energy effective theory of heterotic string theory, whereas the Kerr solution is described only by mass and angular momentum and is formulated within GR.

 \cite{chandrasekhar1998mathematical, uniyal2017null, garfinkle1992erratum}.\\

\noindent
In order to test modified gravity, to reveal possible deviations from GR, and to understand its effects on astrophysical phenomena, it is important to understand the behavior of matter around black holes. The accretion flow around black holes is particularly important because the properties of the spacetime directly affect the morphology of the accreting matter falling toward the black hole and allow the formation of a physical mechanism. At the same time, these spacetime properties cause changes in the variability of the matter accreting toward the black hole and in the properties of the QPOs observed in X-ray binary systems and active galactic nuclei (AGN) \cite{Bambi:2015kza}. Thus, hydrodynamical and timing diagnostics provide important ways to establish a connection between modified-gravity black holes and observed astrophysical sources.\\

 \noindent
In order to test modified-gravity black-hole solutions and to reveal their possible deviations from the Kerr spacetime, modeling the behavior of matter accreting around black holes through the BHL mechanism in the strong gravitational field may provide an important way to understand these effects. In this context, the BHL accretion mechanism provides a useful physical setup because it allows the gas to fall supersonically toward the black hole and leads to the formation of a shock cone on the downstream side of the accretion flow \cite{Bondimnras_104.5.273, 1939PCPS...34..405H}. In relativistic spacetime geometries, the morphology of this shock cone is directly affected by the near-horizon geometry of the black hole, the black-hole spin, and the additional deformation parameters that characterize the spacetime \cite{Font:1999ms, Penner:2010px, Lora-Clavijo:2015hqa, Donmez2024MPLA, Donmez2025JHEAp, donmez2026accretion, donmez2026relativistic}. Therefore, by examining the density distribution around the black hole, the position of the shock cone, and the time-dependent behavior of the mass-accretion rate, the hydrodynamical signatures of the underlying spacetime can be revealed, and the effects of modified gravity on the accretion dynamics in the strong-gravity region can be understood.\\

 \noindent
The shock cone formed around the black hole through the BHL mechanism can behave as a dynamical cavity, allowing the fundamental oscillation modes to be trapped and excited inside this cavity. These trapped modes may lead to the formation of QPO-like signatures. In the strong gravitational field, the oscillation frequencies can be extracted by performing PSD analyses of the numerically computed mass-accretion-rate signal \cite{Chakrabarti:2002ur, Donmez2012MNRAS, Donmez2024JCAP, Donmez:2026yvm}. Since QPOs are commonly observed in X-ray binary systems and AGNs, they provide direct timing tools for comparing observational data with numerical results \cite{Motta:2013wga, Ingram:2014ara, Remillard:2006fc, Motta:2016vwf, Ingram:2019mna}. Thus, the characteristic frequencies obtained from relativistic hydrodynamical simulations can be used to establish a connection between the shock-cone dynamics formed around modified-gravity black holes and the observed QPOs, harmonic ratios, and mass-scaled temporal signatures. \\

\noindent
The QPO-like behavior obtained numerically can be used to test gravity by comparing it with observed sources across different black-hole mass scales. In stellar-mass black-hole systems, sources such as GRS~1915+105 \cite{Belloni:2013qka, Belloni:2019sot, Sreehari:2020jge} and IGR~J17091--3624 \cite{Iyer:2015tga, Debnath:2025sep} have been observed to produce high-frequency QPOs in the tens of Hz range. For GRS~1915+105, the best-known QPO pair appears at approximately $41$ and $67~\mathrm{Hz}$, while for IGR~J17091--3624, a QPO has been reported at around $66~\mathrm{Hz}$. In the intermediate-mass black-hole regime, the ultraluminous X-ray sources M82~X--1 and NGC~5408~X--1 appear as important observational targets for comparison. In M82~X--1, twin QPO peaks have been observed at $3.32$ and $5.07~\mathrm{Hz}$ \cite{Pasham:2014ybe, Mucciarelli:2005jx}. On the other hand, for NGC~5408~X--1, the observed QPOs occur around $0.010$--$0.020~\mathrm{Hz}$ \cite{Pasham:2012thw}. In supermassive black-hole systems associated with AGNs, some of the observed QPO sources are RE~J1034+396 \cite{2010MNRAS.401..507B}, 1H~0707--495 \cite{Fabian2012}, and ESO~113--G010 \cite{2020ChAA..44...32Z}. The QPO frequencies observed from these sources are generally in the range of approximately $10^{-5}$--$10^{-4}~\mathrm{Hz}$. Therefore, by applying inverse mass scaling to the characteristic frequencies obtained from relativistic hydrodynamical simulations, the same shock-cone oscillation mechanism can be compared with observed QPOs from stellar-mass, intermediate-mass, and supermassive black-hole systems. This comparison allows us to examine whether Kerr--Sen spacetime effects can provide a unified timing interpretation over a wide range of black-hole masses. \\

\noindent
In this paper, we numerically solve the GRH equations for BHL accretion around the Kerr and Kerr--Sen black holes in order to reveal how the charge-related deformation modifies the Kerr--Sen spacetime and how these effects appear in the strong gravitational field. By comparing the results obtained for the Kerr--Sen black-hole models with the corresponding Kerr solutions, we provide a framework for testing the Kerr--Sen geometry against the standard Kerr spacetime. In particular, we investigate how the Kerr--Sen deformation affects the morphology of the shock cone formed around the black hole, the density distribution, the position of the shock, and the variations in the mass-accretion rate. In this way, we identify the hydrodynamical deviations produced by the additional spacetime parameter. Then, in order to reveal the oscillatory properties of the mass-accretion-rate signal computed in the strong gravitational field and to discuss the excitation and observability of the fundamental modes trapped inside the shock cone, we perform PSD analysis and apply multi-component Lorentzian fits to extract the dominant QPO-like oscillation modes. Finally, by applying inverse mass scaling to the numerically obtained characteristic frequencies, we compare the results found in this paper with the observed QPOs from stellar-mass, intermediate-mass, and supermassive black-hole systems. Thus, we examine whether Kerr--Sen shock-cone oscillations can provide a physical mechanism for the observed timing signatures. At the same time, numerically calculated QPOs allow us to make predictions for physical parameters, such as the mass and spin, of observed sources whose properties are not fully determined observationally.\\

\noindent
The plan of the paper is as follows. In Section~\ref{ferame}, we present the numerical framework used in this study. In this section, we describe the general relativistic hydrodynamics equations, the Kerr--Sen spacetime, the horizon structure, and the numerical setup adopted for the BHL accretion simulations. In Section~\ref{Num1}, we analyze the shock-cone morphology, the variation of the mass-accretion rate, and the QPO-like oscillation modes obtained from the PSD and Lorentzian-fit analyses of the numerical simulations. In this way, we reveal the hydrodynamical and temporal signatures of the deviations of the Kerr--Sen spacetime from the Kerr spacetime. In Section~\ref{compare1}, we compare the characteristic frequencies obtained from the numerical calculations with the observed QPOs from stellar-mass, intermediate-mass, and supermassive black-hole systems. Finally, in Section~\ref{conc}, we summarize the main results and discuss their implications for identifying possible Kerr--Sen signatures in astrophysical black-hole observations. Unless otherwise stated, we use geometrized units with $M=c=G=1$ throughout this paper.

 \section{Numerical Framework and Kerr–Sen Spacetime}
 \label{ferame}

 \subsection{General Relativistic Hydrodynamics Equations and Kerr--Sen Spacetime}
 \label{GRH_metric}
 
In order to reveal the accretion morphology produced by mass accretion around black holes and to identify the oscillation modes trapped or excited as a result of the interaction between the black hole and the accreting matter, we numerically solve the GRH equations. The covariant form of the GRH equations and the derivation of their conservative form have been discussed in detail in Refs.~\cite{Font2000NumericalHI, Donmez2004ASS}. The GRH equations in conservative form are written as \cite{Font2000NumericalHI, Donmez2004ASS, Donmez2006AMC, Donmez2012MNRAS} 
\begin{equation} 
\frac{\partial \mathbf{U}}{\partial t} + \frac{\partial \mathbf{F}^{i}}{\partial q^{i}} = \mathbf{S}, 
\label{eq:grh_conservative} 
\end{equation} 
\noindent 
where $\mathbf{U}$, $\mathbf{F}^{i}$, and $\mathbf{S}$ are the vectors of the conserved variables, fluxes, and source terms, respectively, while $q^{i}$ denotes the generalized coordinate in the $i$th direction. Since the numerical simulations in this study are performed in two dimensions on the equatorial plane, Eq.~\ref{eq:grh_conservative} can be rewritten as 
\begin{equation} 
\frac{\partial \mathbf{U}}{\partial t} + \frac{\partial \mathbf{F}^{r}}{\partial r} + \frac{\partial \mathbf{F}^{\phi}}{\partial \phi} = \mathbf{S}. 
\label{eq:grh_2d}
 \end{equation}

\noindent
Here, $\mathbf{U}$ represents the vector of conserved variables and can be written in terms of the primitive variables, the metric functions of the spacetime, and the thermodynamical quantities as follows:

\begin{equation}
\mathbf{U} = \begin{pmatrix} D \\ S_{r} \\ S_{\phi} \\ \tau \end{pmatrix} = \begin{pmatrix} \sqrt{\gamma} W \rho \\ \sqrt{\gamma} \rho h W^{2} V_{r} \\ \sqrt{\gamma} \rho h W^{2} V_{\phi} \\ \sqrt{\gamma}\left(\rho h W^{2} - p - W\rho\right) \end{pmatrix}. 
\label{eq:conserved_variables} 
\end{equation}

\noindent 
Here, $V^{i}$ is the three-velocity of the fluid, $W$ is the Lorentz factor, $\gamma$ is the determinant of the spatial three-metric, $h$ is the specific enthalpy, $\rho$ is the rest-mass density, and $p$ is the pressure. The flux vectors appearing in Eq.~\ref{eq:grh_2d} are written as
 
\begin{equation}
\mathbf{F}^{r} = \begin{pmatrix} \alpha\left(V^{r}-\dfrac{\beta^{r}}{\alpha}\right)D \\ \alpha\left[\left(V^{r}-\dfrac{\beta^{r}}{\alpha}\right)S_{r} +\sqrt{\gamma}\,p\right] \\ \alpha\left(V^{r}-\dfrac{\beta^{r}}{\alpha}\right)S_{\phi} \\ \alpha\left[\left(V^{r}-\dfrac{\beta^{r}}{\alpha}\right)\tau +\sqrt{\gamma}\,V^{r}p\right] \end{pmatrix},
\label{eq:flux_r} 
\end{equation} 

\begin{equation} 
\mathbf{F}^{\phi} = \begin{pmatrix} \alpha\left(V^{\phi}-\dfrac{\beta^{\phi}}{\alpha}\right)D \\ \alpha\left(V^{\phi}-\dfrac{\beta^{\phi}}{\alpha}\right)S_{r} \\ \alpha\left[\left(V^{\phi}-\dfrac{\beta^{\phi}}{\alpha}\right)S_{\phi} +\sqrt{\gamma}\,p\right] \\ \alpha\left[\left(V^{\phi}-\dfrac{\beta^{\phi}}{\alpha}\right)\tau +\sqrt{\gamma}\,V^{\phi}p\right] \end{pmatrix}. 
\label{eq:flux_phi} 
\end{equation}

\noindent 
And the source term is

\begin{equation} 
\mathbf{S} = \begin{pmatrix} 0 \\ \alpha\sqrt{\gamma}\,T^{\mu\nu}g_{\nu\sigma}\Gamma^{\sigma}_{\mu r} \\ \alpha\sqrt{\gamma}\,T^{\mu\nu}g_{\nu\sigma}\Gamma^{\sigma}_{\mu \phi} \\ \alpha\sqrt{\gamma} \left[ T^{rt}\partial_{r}\alpha - \alpha \left( T^{rr}\Gamma^{t}_{rr} + T^{r\phi}\Gamma^{t}_{r\phi} \right) \right] \end{pmatrix}. 
\label{eq:source_vector} 
\end{equation}

\noindent
In this study, the ideal-gas equation of state given below is used for the system: 
\begin{equation}
p = (\Gamma - 1)\rho \epsilon, 
\label{eq:eos} 
\end{equation} 
\noindent
where $\Gamma$ is the adiabatic index, $\rho$ is the rest-mass density, and $\epsilon$ is the specific internal energy. The adiabatic index is chosen as $\Gamma=4/3$, which is appropriate for describing a relativistic gas flow in the strong-gravity region around black holes. We solve Eq.~\ref{eq:grh_2d} by using high-resolution shock-capturing methods, which are well suited for accurately resolving discontinuities and shock structures in relativistic accretion flows. The numerical fluxes are computed using the Marquina flux formula, while the primitive variables are reconstructed at cell interfaces by using the MUSCL scheme. This combination allows the shock cone and the steep gradients formed in the downstream region to be resolved without introducing excessive numerical diffusion. The details of the numerical solution of these equations, the different numerical methods employed, the test problems, and the corresponding convergence tests have been presented in detail in our previous studies  \cite{Donmez2004ASS, Donmez2006AMC, Donmez2012MNRAS} .

\noindent
As seen from Eqs.~\ref{eq:conserved_variables}--\ref{eq:source_vector} , the GRH equations explicitly depend on the spacetime metric. Since in this study we describe the accretion dynamics around a Kerr--Sen black hole in a fixed background spacetime, we adopt the Kerr--Sen metric. In Boyer--Lindquist coordinates $(t,r,\theta,\phi)$, the line element of the Kerr--Sen spacetime is written as \cite{uniyal2017null,siahaan2016destroying,ghosh2021parameters,zhang2021escape,izmailov2020string,sakti2023hidden}
 
 \begin{align}
 ds^{2}&&=-\left(\frac{\Delta- a^{2}\sin^{2}\theta}{\rho^{2}}\right) dt^{2} -\left(\frac{4Mra\sin^{2}\theta}{\rho^{2}}\right) d\phi dt+\frac{\rho^{2}}{\Delta} dr^{2}\nonumber\\&&+\rho^{2} d\theta^{2} +\left(\rho^{2}+a^{2}\sin^{2}\theta+\frac{2Mra^{2}\sin^{2}\theta}{\rho^{2}}\right)\sin^{2}\theta d\phi^{2},
 \end{align}
 where,~~ $\rho^{2}=r\left(r+q\right)+a^{2}\cos^{2}\theta$~~ and ~$\Delta=r\left(r+q\right)-2Mr+a^{2}$. \\ Here, $q$ stands for $\frac{Q^{2}}{M}$; where, $Q$ and $M$ represent the charge and mass of the black hole, respectively, and $a=J/M$ represent the  dimentionless black hole spin with angular momentum $J$.\\

\begin{figure*}[!t]
\centering
\includegraphics[width=0.48\textwidth]{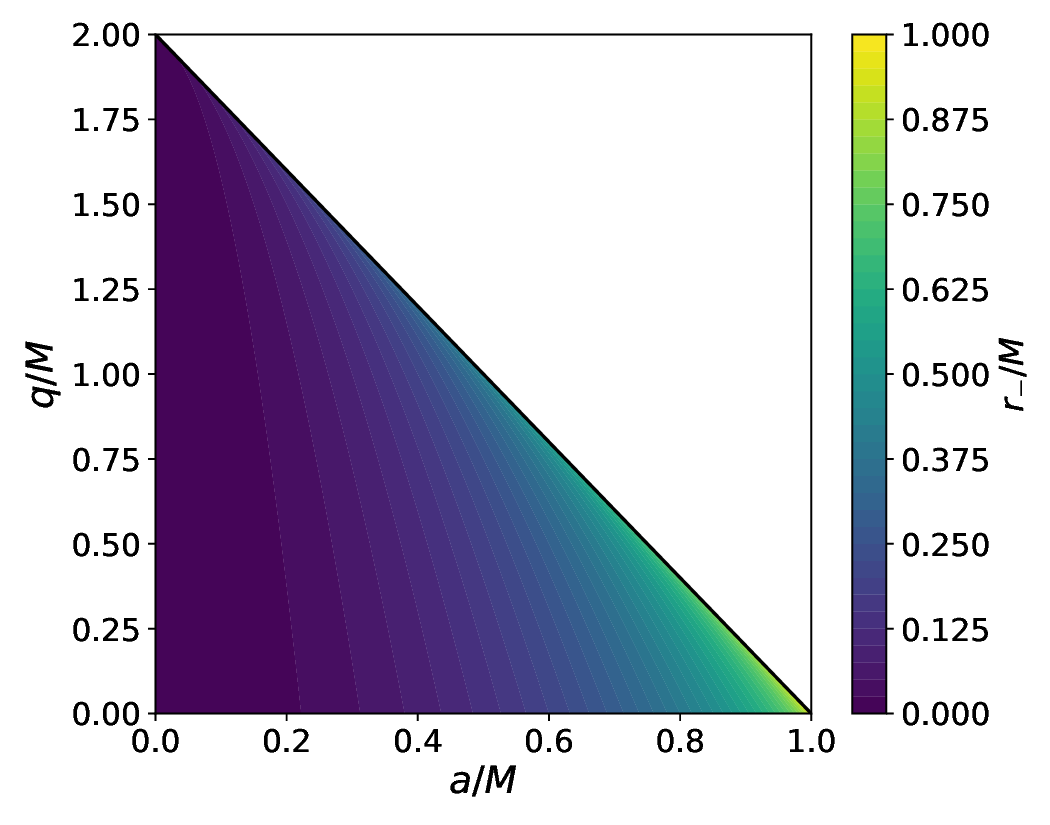}
\includegraphics[width=0.48\textwidth]{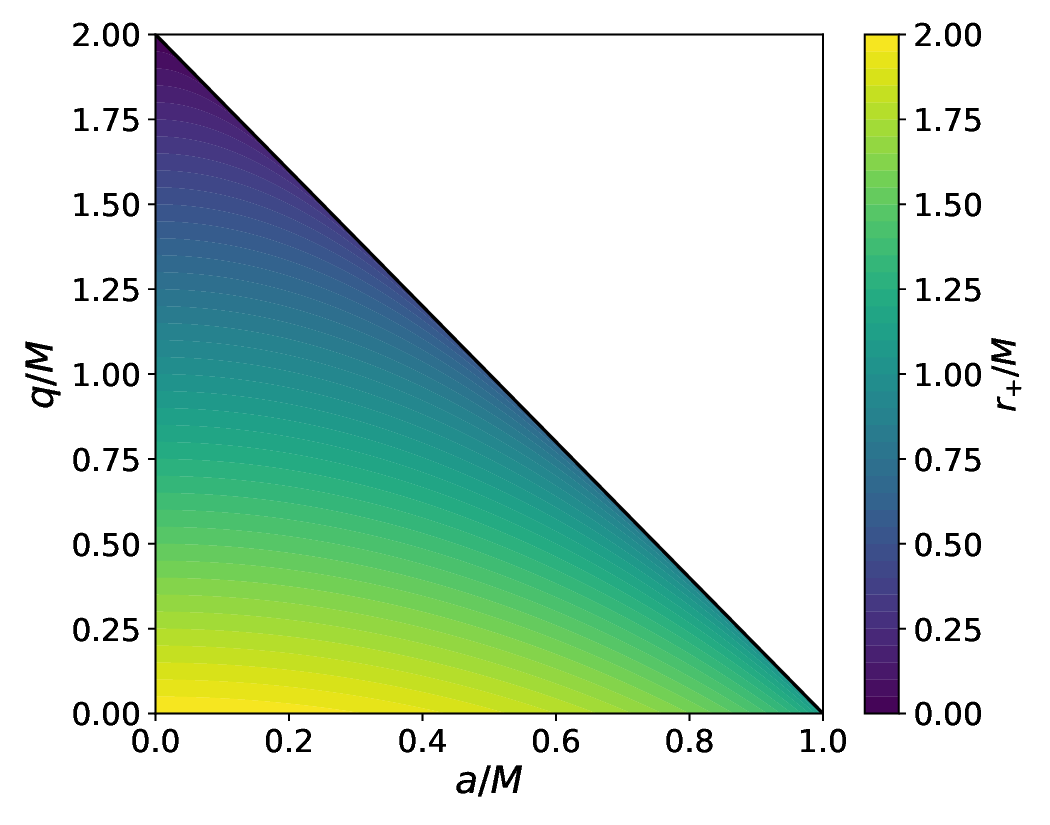}
\caption{The inner and outer horizons of the Kerr--Sen black hole in the $(a/M, q/M)$ parameter plane are shown for $M=1$. The left panel displays the inner horizon radius $r_{-}/M$, while the right panel shows the outer horizon radius $r_{+}/M$. The colored region represents the parameter space in which a classical black hole exists, whereas the white region corresponds to the formation of a naked singularity.}
\label{horizon}
\end{figure*}

\noindent
The horizon structure of the Kerr--Sen spacetime is defined by the condition $\Delta=0$. Thus, the existence of real and positive horizon radii depends directly on the spin parameter of the black hole and the charge-related deformation parameter. In Fig.~\ref{horizon}, the possible classical black-hole region in the $a/M$--$q/M$ plane is shown for the case $M=1$. The colored region represents the parameter values that produce horizons in the Kerr--Sen solution. Therefore, this colored region corresponds to the black-hole spacetime. On the other hand, the white region shows the parameter range in which a classical black hole does not exist and a naked singularity is formed. The left panel shows the inner horizon, corresponding to the Cauchy horizon, while the right panel shows the outer horizon, corresponding to the event horizon. As the Kerr--Sen deformation parameter increases, the possible black-hole formation region becomes progressively narrower. This shows that the charge-related deformation strongly affects the causal structure of the spacetime. The parameters used in this study, which are given in Table~\ref{tab:KS_models}, are entirely selected from the colored region in Fig.~\ref{horizon}. The spin parameter, deformation parameter, and corresponding event-horizon radius of each Kerr--Sen black-hole model are listed in Table~\ref{tab:KS_models}. Therefore, all models used in these numerical simulations produce black-hole solutions.

 \subsection{Numerical Setup and Approximations}
 \label{Num_setup}
The numerical simulations are performed on the equatorial plane of the Kerr and Kerr--Sen black-hole spacetimes. Throughout the simulations, the black-hole mass is set to $M=1$, and geometrized units are used. The spacetime is treated as a fixed background, and the evolution is carried out only for the relativistic hydrodynamical variables. In order to inject matter toward the black hole from the upstream region at the outer boundary of the computational domain, the initial velocity field is prescribed in terms of an asymptotic velocity $V_{\infty}$ following the standard BHL configuration. Following the standard relativistic BHL prescription used in previous studies \cite{Font:1999ms,Penner:2010px, Lora-Clavijo:2015hqa, Donmez2024MPLA, Donmez2025JHEAp, Donmez:2025piv}, the radial and angular components of the initial three-velocity are written as

\begin{equation} V^{r}=\sqrt{\gamma^{rr}}\,V_{\infty}\cos\phi, \label{eq:vr_initial} 
\end{equation} 
and 
\begin{equation} V^{\phi}=-\sqrt{\gamma^{\phi\phi}}\,V_{\infty}\sin\phi . \label{eq:vphi_initial} 
\end{equation} 
Here, $V^{r}$ and $V^{\phi}$ are the radial and angular components of the initial three-velocity, respectively, while $\gamma^{rr}$ and $\gamma^{\phi\phi}$ are the corresponding contravariant components of the spatial metric. These relations define an initially uniform flow at large distance, with the gas injected parallel to the upstream direction. 
In this study, the asymptotic velocity of the gas is chosen as $V_{\infty}=0.2$. The asymptotic sound speed is set to $c_{s,\infty}=0.1$. Therefore, the corresponding Mach number is $\mathcal{M}=2$, showing that the matter falling toward the black hole is injected supersonically. This supersonic motion leads to the formation of a well-defined shock cone in the downstream region and provides a suitable setup for studying the response of the accretion flow to the Kerr--Sen deformation in the strong-gravity region.\\

\noindent
At the outer boundary of the computational domain, an inflow boundary condition is imposed in the upstream region in order to continuously inject matter toward the black hole, while outflow boundary conditions are applied in the remaining outer-boundary regions. In this way, matter leaving the computational domain from the downstream region is removed without producing artificial reflections. At the inner boundary close to the black-hole horizon, an outflow boundary condition is also used so that matter can freely fall into the black hole. Periodic boundary conditions are imposed in the angular direction. The mass-accretion rate is measured near this inner boundary, where the flow carries the imprint of the shock-cone dynamics and the near-horizon structure of the spacetime. 
The simulations are carried out on a two-dimensional computational grid with $1024$ grid points in the radial direction and $512$ grid points in the angular direction.  This resolution allows us to resolve the shock-cone morphology, the density variation inside the cone, and the time-dependent variation of the mass-accretion rate used in the PSD analysis.  The computational domain extends from $r_{\min}=2.3M$ to $r_{\max}=100M$. The inner radial boundary is therefore located outside the event horizon for all Kerr and Kerr--Sen models considered in this work. In particular, the largest event-horizon radius among the models listed in Table~ \ref {tab:KS_models} is smaller than $r_{\min}$, ensuring that the inner boundary is placed consistently outside the corresponding horizons.  The spacetime parameters of the Kerr and Kerr--Sen models used in this paper are listed in Table~\ref{tab:KS_models}. The Kerr models are used as reference solutions, while the Kerr--Sen models are constructed by varying the charge-related deformation parameter $q$. For each model, the corresponding event-horizon radius is also given in Table~I, ensuring that all simulations are performed for black-hole solutions and that the inner boundary is chosen consistently with the causal structure of each spacetime.

\begin{table*}[!t]
\centering
\caption{Set of Kerr--Sen black hole models adopted as initial backgrounds for the BHL accretion simulations with $M=1$. In the table, $a$ denotes the black hole spin parameter, $q=Q^{2}/M$ is the charge-related deformation parameter characterizing the deviation from the Kerr spacetime, and $r_{+}$ represents the radius of the outer event horizon. According to the magnitude of $q$, the models are classified into soft, moderate, and strong modification regimes.}
\label{tab:KS_models}
\setlength{\tabcolsep}{12pt}
\begin{tabular}{lcccc}
\hline\hline
Model & $a[M]$ & Modification class & $q[M]$ & $r_{+}[M]$ \\
\hline
Kerr & 0.9 & --       & 0.0 & 1.4359 \\
KS1  & 0.9 & Soft     & 0.01 & 1.419 \\
KS2  & 0.9 & Moderate & 0.08 & 1.294 \\
KS3  & 0.9 & Strong   & 0.16 & 1.111 \\
KS4  & 0.9 & Strong   & 0.19 & 1.000 \\
\hline
Kerr & 0.5 & --       & 0.0 & 1.866 \\
KS5  & 0.5 & Moderate & 0.30 & 1.537 \\
KS6 & 0.5 & Strong   & 0.90 & 0.779 \\ 
\hline\hline
\end{tabular}
\end{table*}

\section{Hydrodynamical and Timing Signatures of Kerr–Sen Deviations from Kerr}
\label{Num1}

In this section, we reveal the hydrodynamical and temporal imprints of the deviations of the Kerr--Sen black hole results from the Kerr spacetime solutions by investigating the accretion dynamics formed around the black hole through BHL accretion. Using the numerical setup described in Section \ref{Num_setup}, we compare the Kerr–Sen models with the corresponding Kerr reference solutions in order to determine how the charge-related deformation parameter modifies the shock-cone morphology, the mass-accretion-rate variability, and the QPO-like oscillation modes in the strong-gravity region. The Kerr solutions presented in this paper have also been used in our previous studies and are given as reference models \cite{Donmez2012MNRAS, Donmez2024JCAP, donmez2026relativistic, donmez2026accretion}. In this way, the effects of the deformation parameter associated with the charge in the Kerr--Sen model, $q=Q^2/M$, especially the changes it produces in the near-horizon region of the black hole, are examined in the strong gravitational field and compared with the Kerr solutions by considering its interaction with the black hole spin parameter. By considering two representative spin families, $a=0.9M$ and $a=0.5M$, we investigate how different levels of deviation change the accretion flow and the resulting morphological structure. In particular, the variations of the rest-mass density in the azimuthal direction, together with the changes and oscillations of the mass-accretion rates, are analyzed comparatively in order to reveal the observable physical consequences in the strong gravitational field. These diagnostics allow us to quantify how the shock-cone structure, density enhancement, and accretion variability differ from the corresponding Kerr cases. In addition, by performing PSD analysis together with Lorentzian fitting of the mass-accretion rate, the characteristic oscillation modes in the strong gravitational field and the details of the observability of these modes are presented. Throughout this study, the Kerr models are presented as reference solutions, and based on these reference solutions, we aim to show the deviation of the Kerr--Sen solutions from the general relativistic Kerr case and to identify the conditions under which their hydrodynamical and temporal signatures can be measured.

\subsection{Shock-Cone Morphology in the Strong-Gravity Region}
\label{Num3}
\begin{figure*}[htbp]
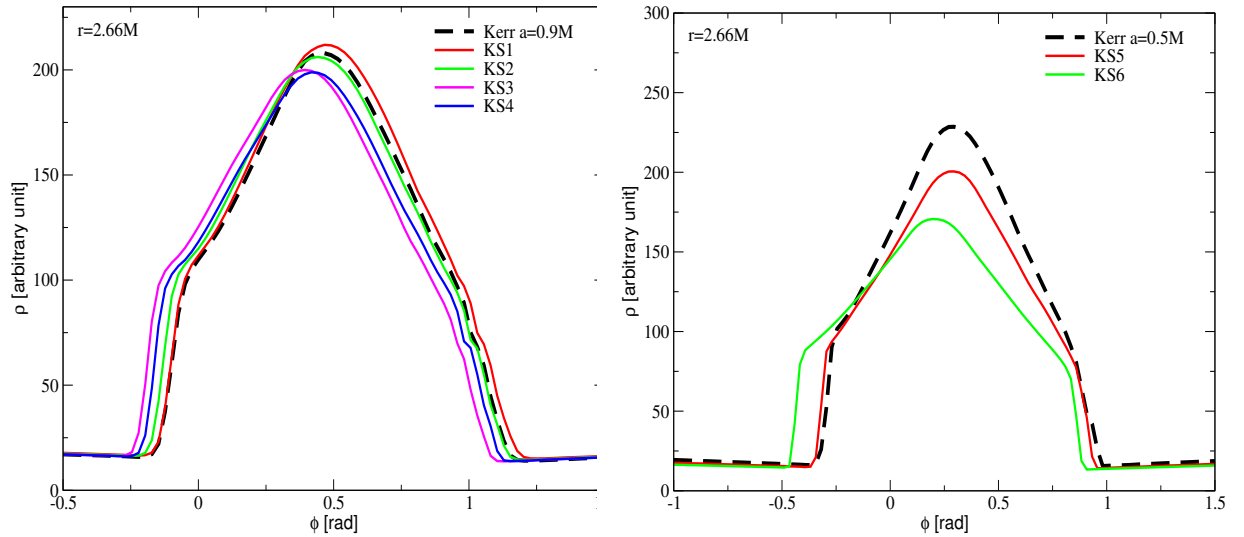

\centering
\includegraphics[width=8.0cm,height=7.0cm]{den_phi_a09_r266.eps}
\includegraphics[width=8.0cm,height=7.0cm]{den_phi_a05_r266.eps}
\caption{Azimuthal variation of the rest-mass density around Kerr and Kerr--Sen black holes at the fixed radius $r=2.66M$. The left panel shows the variations for the spin parameter $a=0.9M$, while the right panel shows the same behavior for the moderate-spin black hole models with $a=0.5M$. When the Kerr and Kerr--Sen models are compared, it is seen that the Kerr--Sen deformation parameter modifies the shock-cone morphology in the strong gravitational field, changes the density of the matter inside the cone, and alters the angular shock structure.}\label{dens_a09a05}
\end{figure*}

The effects of the Kerr--Sen spacetime metric on the dynamics of the shock cone formed in the downstream region may affect the fundamental modes trapped inside the shock cone and their propagation, as well as the amplitude of the oscillations. For this reason, in Fig.~\ref{dens_a09a05}, we show how the azimuthal distribution of the rest-mass density of the shock cone formed around Kerr and Kerr--Sen black holes changes at $r=2.66M$ in the strong gravitational field. As seen in Fig.~\ref{dens_a09a05}, the Kerr--Sen geometry significantly affects the shock cone formed in the strong gravitational region. Thus, this figure shows that the Kerr--Sen deformation can influence the physical mechanisms occurring around the black hole and, accordingly, the related physical phenomena. Since the density is measured in the azimuthal direction, the width of the high-density region indicates how large an angular region is covered by the shock cone. On the other hand, the peak value in the density profile provides information about the matter compressed inside the cone. Therefore, the change in the position of the shock cone, the change in its width, and the change in the maximum peak value can be interpreted as hydrodynamical signatures of the modified gravitational field.

In the left panel of Fig.~\ref{dens_a09a05}, the results are presented comparatively for the rapidly rotating black hole models with $a=0.9M$, including the Kerr reference model and the Kerr--Sen models. As seen in both the Kerr and Kerr--Sen solutions, the maximum peaks of the shock cone occur around $\phi=0.5~\mathrm{rad}$. In non-rotating black hole models, the maximum peaks are normally observed around $\phi=0$.  Therefore, the appearance of these peaks in the azimuthal direction is entirely due to the effect of the spin parameter in the strong gravitational field through the bending of spacetime. On the other hand, in the case of $a=0.9M$, when compared with the behavior of the $a=0.5M$ models shown in the right panel of Fig.~\ref{dens_a09a05}, the Kerr--Sen profiles exhibit a relatively close behavior to the Kerr solution. The differences between these models arise entirely as a result of the effect of the spacetime deformation parameter $q$. In all models shown in the left panel of Fig.~\ref{dens_a09a05}, a broad high-density region is formed in the downstream region. At the boundaries of the cone, the density gradient is very steep. Systematic differences appear among the models in the angular width and in the maximum density inside the cone. The shock cones formed in the Kerr--Sen models, together with the shock locations and the values of the density peaks, are slightly modified. This implies that the deformation parameter $q$ changes the gravitational focusing of the matter falling toward the black hole. Therefore, these changes indicate that the Kerr--Sen geometry causes measurable modifications in the structure of the shock cone formed by the physical mechanism around the black hole.

In the right panel of Fig.~\ref{dens_a09a05}, the changes in the dynamical structure of the shock cone formed around Kerr and Kerr--Sen black holes are shown for the lower spin parameter case, $a=0.5M$. In this case, the effect of the Kerr--Sen deformation parameter becomes more pronounced. When compared with the Kerr solution, the Kerr--Sen models show a significant decrease in the density of the matter trapped inside the shock cone. In particular, this decrease is observed more clearly in the highly deformed KS6 model. The maximum density of the cone is considerably lower than that of the Kerr model. In addition, the angular structure of the cone is also significantly modified. The density profile changes its slope near the cone boundaries, and the high-density region becomes less compressed because a wider angular spreading is observed. This suggests that the modified spacetime affects not only the amount of matter trapped inside the cone, but also the angular distribution of the shocked gas.

The behaviors obtained for the two different black hole spin cases shown in Fig.~\ref{dens_a09a05} reveal that the black hole spin plays a significant role in the visibility of the Kerr--Sen deviation in the shock-cone morphology. In the case of $a=0.9M$, the effect of the strong spin parameter and frame dragging may dominate the flow dynamics to some extent and may suppress the relative hydrodynamical imprint of the deformation parameter. As a result, the density profiles obtained in the Kerr--Sen case may produce results similar to those obtained in the Kerr spacetime. On the other hand, in the case of $a=0.5M$, the rotational effect is weaker, and the effect of the Kerr--Sen deformation parameter appears more noticeably in the density distributions. This allows the deformation-induced modification of the gravitational field to produce a stronger change in the shock-cone structure, particularly in the density enhancement inside the cone.

The change in the shock-cone morphology seen in Fig.~\ref{dens_a09a05} can directly affect the oscillations in the behavior of the accretion flow. This is because the shock cone behaves like a cavity around the black hole. The change in the angular width of this cavity and the change in the density of the matter inside the cavity cause changes in the physical properties of the modes trapped and excited inside the cone. While a narrow and dense cone supports stronger mode confinement, broader structures in which less matter is trapped inside the cone change the propagation of perturbations and reduce or shift the dominant oscillatory response. In the Kerr--Sen model, since the deformation parameter of the spacetime changes the amount of matter trapped inside the shock cone and modifies the angular position of the shock cone, the oscillation frequencies of the radial and angular modes excited inside the cone, together with their nonlinear couplings, may be affected by these changes. These hydrodynamical changes provide the physical basis for possible differences in the QPO-like signatures extracted from the mass-accretion-rate variability. These situations are discussed in detail in Sections~\ref{Num4} and \ref{Num5}.

\subsection{Mass-Accretion Variability near the Inner Boundary}
\label{Num4}
The mass-accretion rate measured at the point closest to the black hole, namely at the inner boundary of the computational domain, directly contains diagnostic information about the time-dependent variations of the accretion flow in the strong gravitational field. Since the inner boundary of the computational domain is located at $r=2.23M$, very close to the horizon, the signal of the accretion rate carries information about the matter that passes through the shock cone and is about to fall toward the black hole. Thus, the temporal behavior of the mass-accretion rate carries the combined imprints of the shock-cone oscillations, the matter accumulated inside the cone, gravitational focusing, and the near-horizon structure of the spacetime on the accretion rate. In this context, the variation of the mass-accretion rate provides information not only about the amount of matter captured by the black hole, but also acts as a physical tracer of the dynamical modes excited in the post-shock region. As will be seen later in Section~\ref{Num5}, these time-dependent oscillations appear as characteristic peaks in the PSD analysis. Therefore, the analysis of the mass-accretion rate establishes a bridge between the hydrodynamical structure of the flow and its possible observable temporal signatures.

\begin{figure*}[htbp]
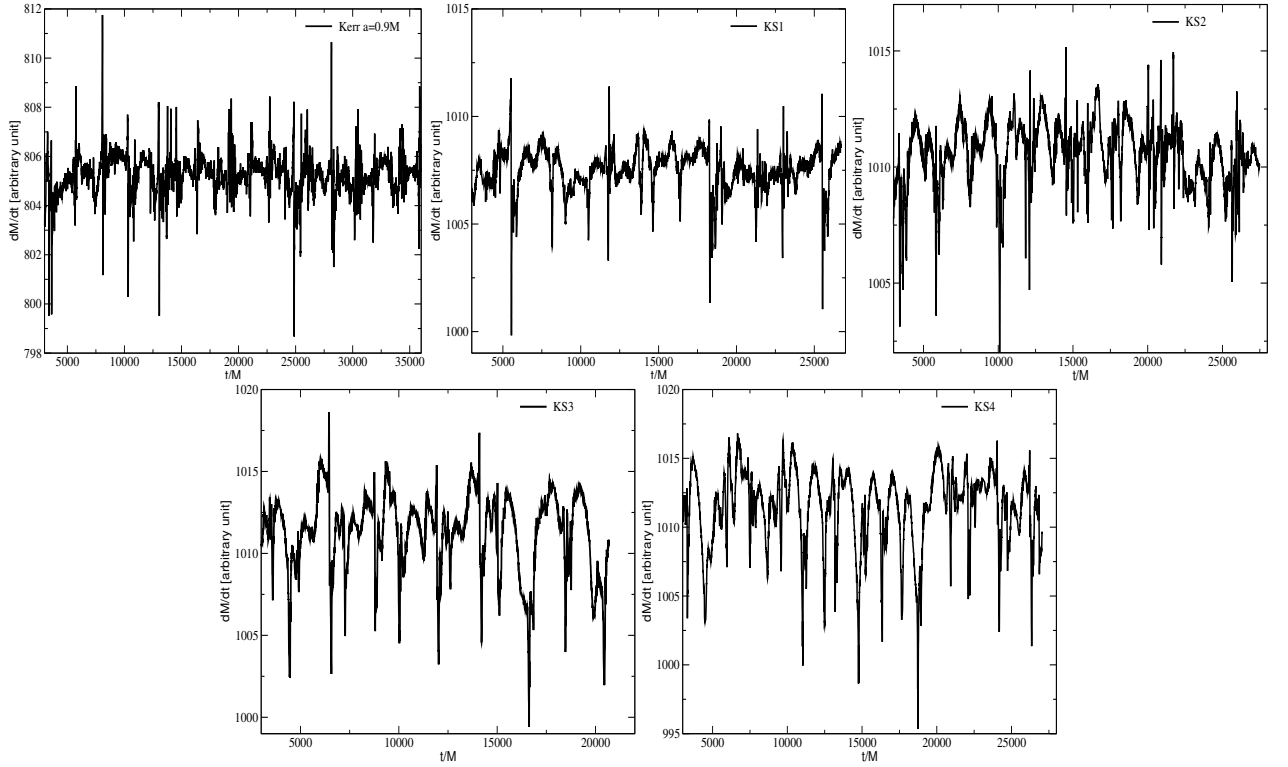

\centering
\includegraphics[width=5.5cm,height=5.0cm]{acc_rate_a09_kerr.eps}
\includegraphics[width=5.5cm,height=5.0cm]{acc_rate_a09_KS1.eps}
\includegraphics[width=5.5cm,height=5.0cm]{acc_rate_a09_KS2.eps}\\
\includegraphics[width=5.5cm,height=5.0cm]{acc_rate_a09_KS3.eps}
\includegraphics[width=5.5cm,height=5.0cm]{acc_rate_a09_KS4.eps}
\caption{Time evolution of the mass-accretion rate at $r=2.33M$ in the strong gravitational field for the rapidly rotating Kerr and Kerr--Sen black hole models with spin parameter $a=0.9M$. The upper-left panel shows the behavior of the mass-accretion rate in the Kerr solution, while the other snapshots show the mass-accretion rates in the Kerr--Sen case for different values of the spacetime deformation parameter $q$. It is clearly seen that the deformation of the spacetime significantly modifies the amount of matter falling into the black hole and the oscillatory behavior of the matter inside the shock cone.}\label{mass_acc_a09}
\end{figure*}

Fig.~\ref {mass_acc_a09} shows the time evolution of the mass-accretion rates computed for the rapidly rotating black hole models with spin parameter $a=0.9M$. The Kerr model shown in the upper-left panel, which is used as the reference model, oscillates within a relatively narrow accretion-rate range around an approximately mean value. At the same time, it exhibits intermittent sharp fluctuations produced by the unsteady motion of the shock cone. The long vertical features in the figure indicate these fluctuations. In the other snapshots of Fig.~\ref {mass_acc_a09}, where the Kerr--Sen deformation starts to modify the spacetime, it is observed that the average mass-accretion rates increase by approximately $20\%$ compared with the Kerr model. This shows that the modified gravity significantly increases the amount of matter falling into the black hole. Among the Kerr--Sen models, KS1 and KS2 produce moderate-level variations at high accretion rates. On the other hand, the KS3 and KS4 models produce relatively stronger and more irregular oscillations. This behavior is especially clearly seen in the KS4 model. As seen in the KS4 model, due to the effect of the strongly deformed spacetime, deeper downward spikes occur in the accretion rate, and more remarkable temporal variations are observed. This shows that the deformation parameter affects not only the stability of the shock cone, but also the efficiency of how much matter is accreted toward the black hole. Although the azimuthal density profiles of the Kerr--Sen models show behavior similar to the Kerr model solution, as seen in Fig.~\ref{dens_a09a05}, the accretion-rate signal shows that the dynamics at the inner boundary are sensitively dependent on the Kerr--Sen deformation.

\begin{figure*}[htbp]
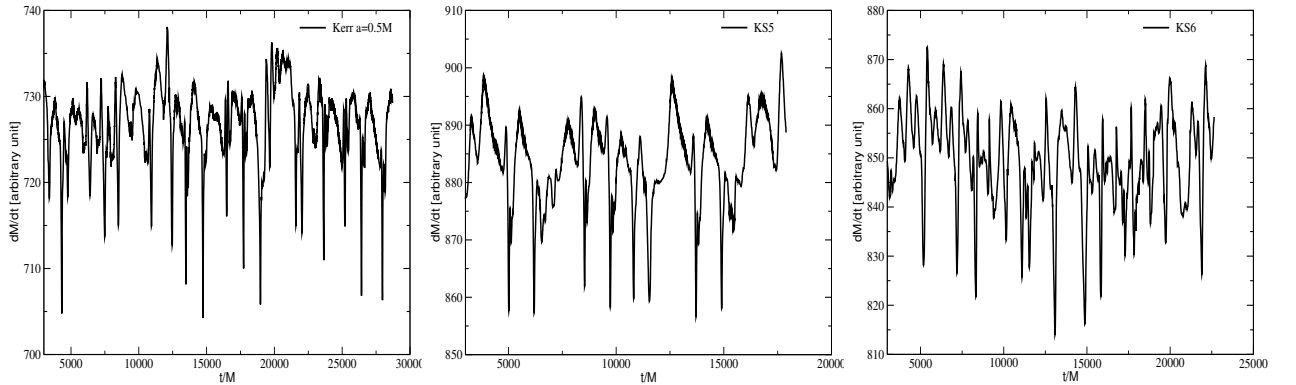

\centering
\includegraphics[width=5.5cm,height=5.0cm]{acc_rate_a05_kerr.eps}
\includegraphics[width=5.5cm,height=5.0cm]{acc_rate_a05_KS5.eps}
\includegraphics[width=5.5cm,height=5.0cm]{acc_rate_a05_KS6.eps}\\
\caption{Same as Fig.~\ref {mass_acc_a09}, but this time the results are shown for the Kerr and Kerr--Sen models with the moderate spin parameter $a=0.5M$. As seen, the deviation in the Kerr--Sen models is significantly stronger than in Fig.~\ref {mass_acc_a09}. This is due to the larger value of the spacetime deformation parameter $q$ and the lower black hole spin parameter compared with Fig.~\ref {mass_acc_a09}.}
\label{mass_acc_a05}
\end{figure*}

Fig.~\ref {mass_acc_a05}  shows the time evolution of the mass-accretion rates for the Kerr and Kerr--Sen models with the moderately rotating black hole spin parameter $a=0.5M$. As seen in Fig.~\ref {mass_acc_a05}, the mean value of the mass-accretion rate in the Kerr solution is lower than those of the Kerr--Sen models. In the Kerr model, recurrent fluctuations associated with the oscillatory behavior of the shock cone are observed. In the Kerr--Sen models, as in the moderately rotating black hole models, the mean value of the mass-accretion rate is higher. This shows that the modified spacetime causes more matter to fall toward the black hole. Among the Kerr--Sen models, KS5 produces strong low-frequency oscillations and large-amplitude variations, indicating that more coherent oscillations may be possible. For the case $a=0.5M$, the KS6 model, which has the strongest deformation, also reveals significant changes in the behavior of the mass-accretion rate. However, in this case, the behavior of the accretion rate is more irregular and spread over a broader band. This means that increasing the deformation applied to the spacetime does not only change the mean mass-accretion rate, but also changes the temporal character of the flow. This temporal character changes depending on how the matter trapped inside the cone is captured, the amount of matter compressed inside it, and the excitation of these modes in the shock-cone region.

When Figs.\ref {mass_acc_a09} and ~\ref {mass_acc_a05}       are compared, it is clearly seen in the accretion dynamics how the black hole spin parameter controls the Kerr--Sen deformation parameter. In the rapidly rotating black hole case, $a=0.9M$, the effects of frame dragging and strong rotation strongly influence the flow. The Kerr--Sen models show accretion-rate variability superposed on a high mean accretion level. In the moderate-spin models, $a=0.5M$, the effect of rotation is weaker, and the effect of the deformation on the spacetime is seen more clearly in the mass-accretion rates. This situation is consistent with the shock-cone morphology given in   Fig.~\ref{dens_a09a05}. Thus, the interaction between the black hole spin and the Kerr--Sen deformation parameters defines both the shock-cone morphology and the temporal behavior of the accretion rate. These results show that the variation in the mass-accretion rate appears as an important condition for observationally distinguishing the Kerr--Sen spacetime from the Kerr solution. These variations reveal the physical reasons for the changes that we observe in the PSD analyses and Lorentzian fits discussed in the next section.

\subsection{QPO-Like Oscillation Modes Induced by Kerr–Sen Deformations}
\label{Num5}
QPO-like oscillation properties can be revealed by performing PSD analysis on time-dependent data. In this study, this analysis is carried out by using the time variation of the mass-accretion rate computed in the strong gravitational field in the previous section. In order to reveal the dominant oscillation components more clearly, the PSD analysis is fitted with multi-component Lorentzian models. This approach is commonly used in black-hole timing studies to characterize QPO peaks and broad-noise components \cite{Nowak:2000ys, Belloni:2002zw}. Each Lorentzian component obtained represents a characteristic mode excited by the accretion flow. This method is useful because the raw PSD contains both narrow and broader coherent peaks. At the same time, the raw PSD may also contain less coherent or non-coherent peaks produced by the nonlinear or artificial oscillations of the shock cone. Numerically separating these cases is important for comparison with observational results and for determining which peaks may be observationally detectable. For example, narrow peaks with large quality factors reveal more coherent oscillation modes and are therefore relevant for observational QPO identification. The quality factor is therefore used as a standard measure of the coherence of QPO-like features in black-hole timing analyses \cite{Remillard:2006fc}. In contrast, broad components reveal less coherent variations distributed over a wide frequency band. Thus, the components obtained by performing Lorentzian fitting provide a quantitative method for separating the characteristic frequencies. At the same time, they allow us to estimate their coherence properties and to compare the temporal behavior of different Kerr--Sen models.

In order to check the robustness of the extracted QPO-like modes, we repeated the PSD analysis by using different portions of the mass-accretion-rate signal and by extracting the signal at an additional radial location in the inner accretion region. The dominant peaks obtained from these tests remain at nearly the same characteristic frequencies, and the same near-harmonic relations are recovered. This shows that the main QPO-like modes reported below are not produced by a particular choice of the time interval or by a single extraction radius. But they represent persistent oscillatory features of the shock-cone dynamics in the strong-gravity region.

\begin{figure*}[htbp]
\centering
\includegraphics[width=8.cm,height=7.0cm]{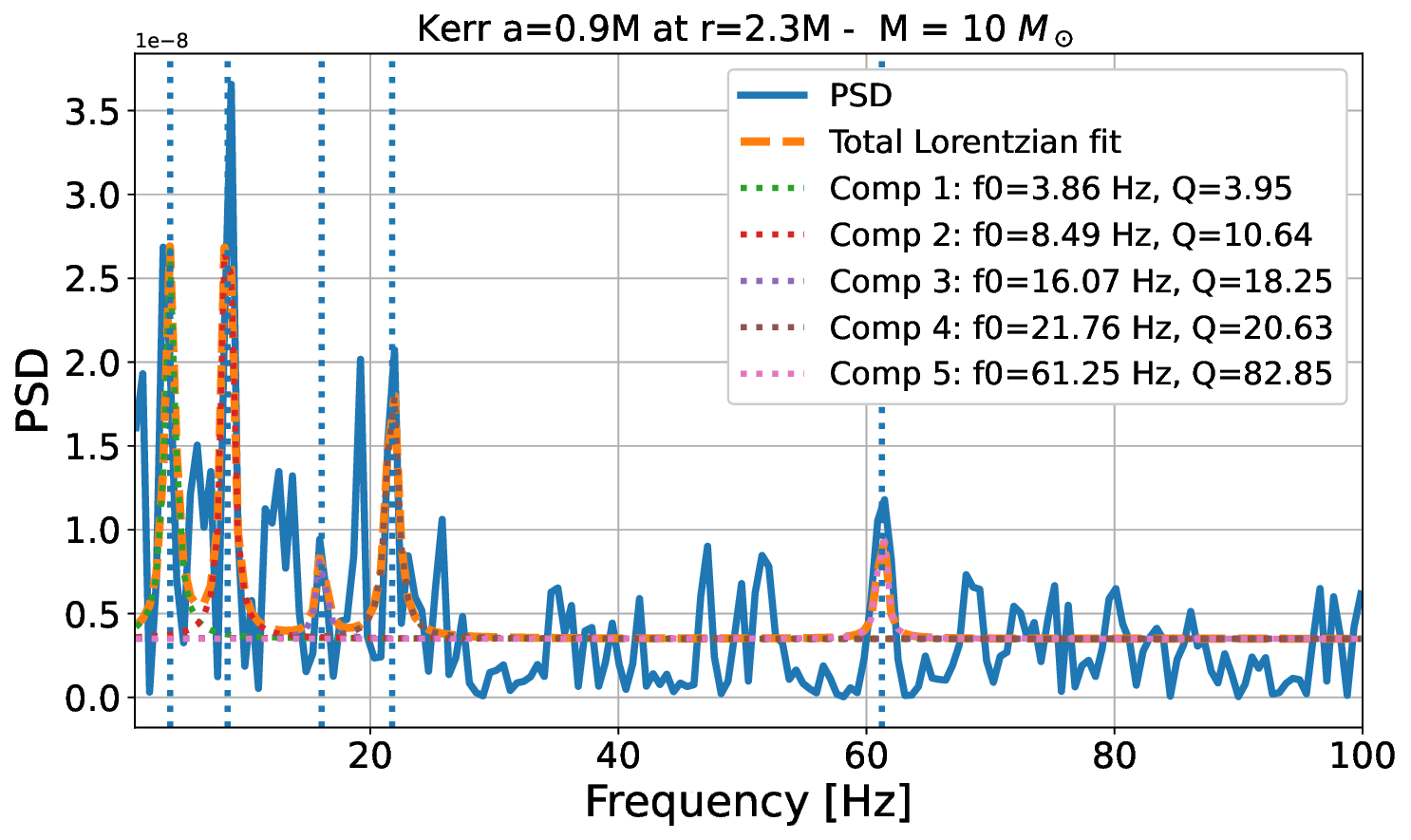}
\includegraphics[width=8.cm,height=7.0cm]{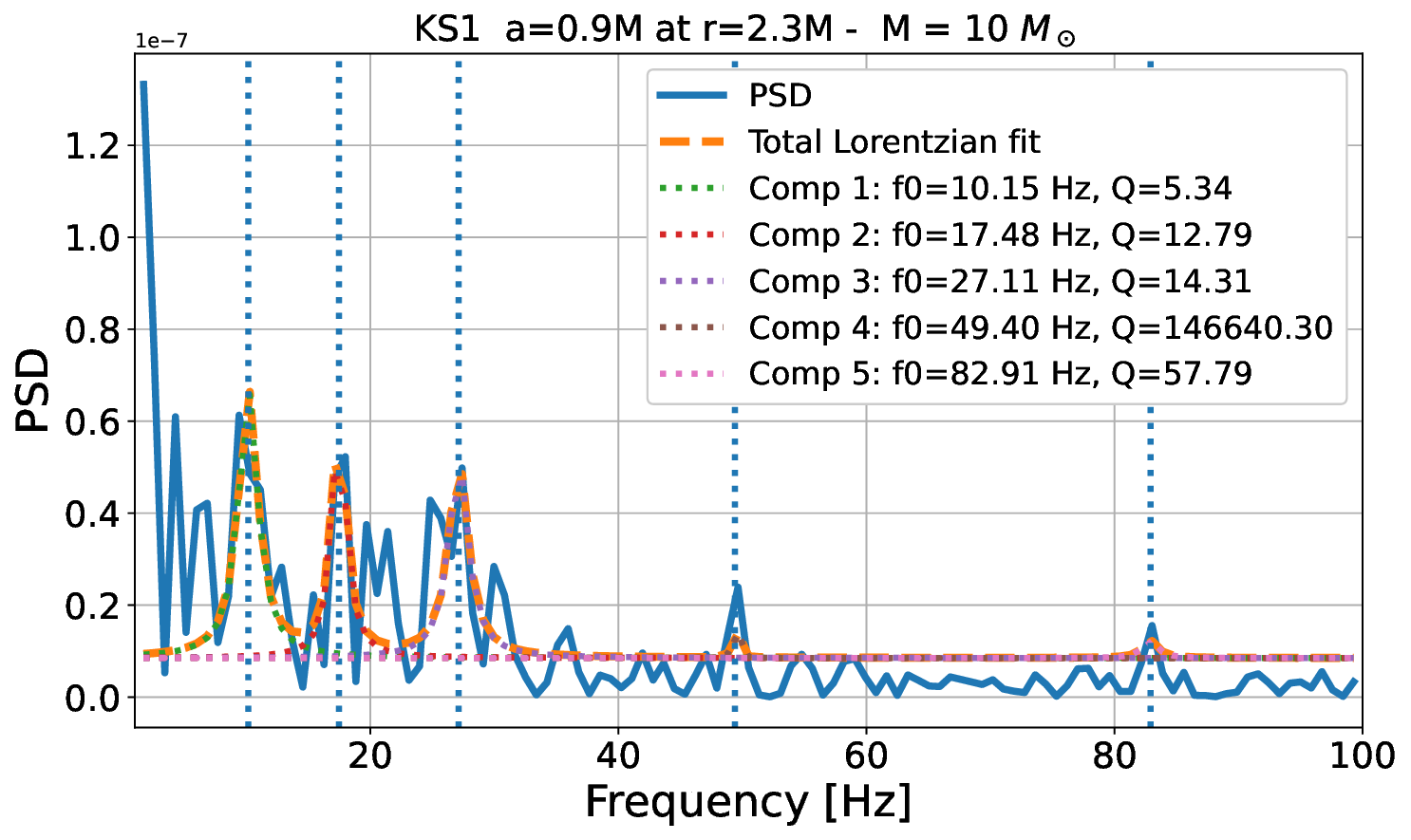} \\
\includegraphics[width=8.cm,height=7.0cm]{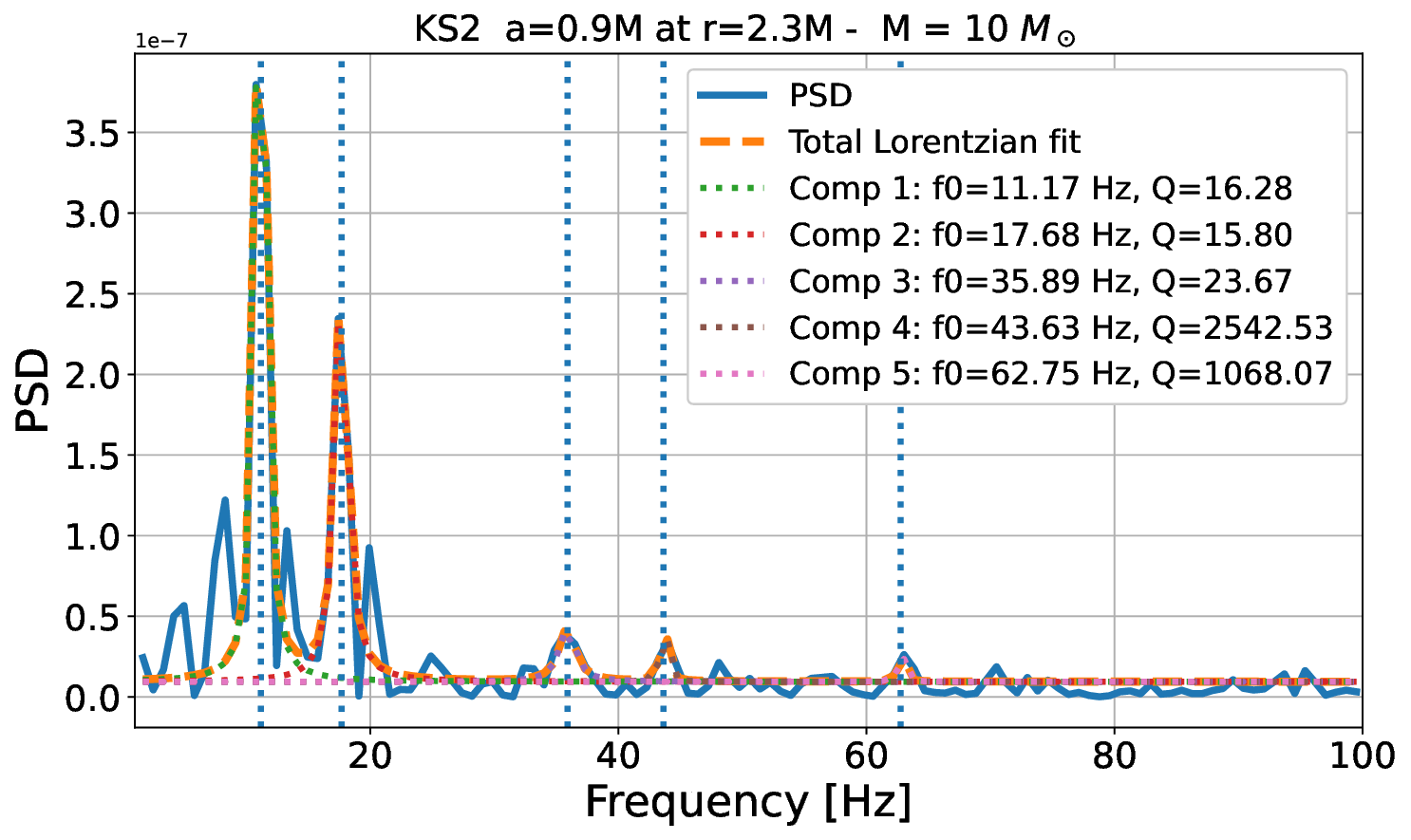}
\includegraphics[width=8.cm,height=7.0cm]{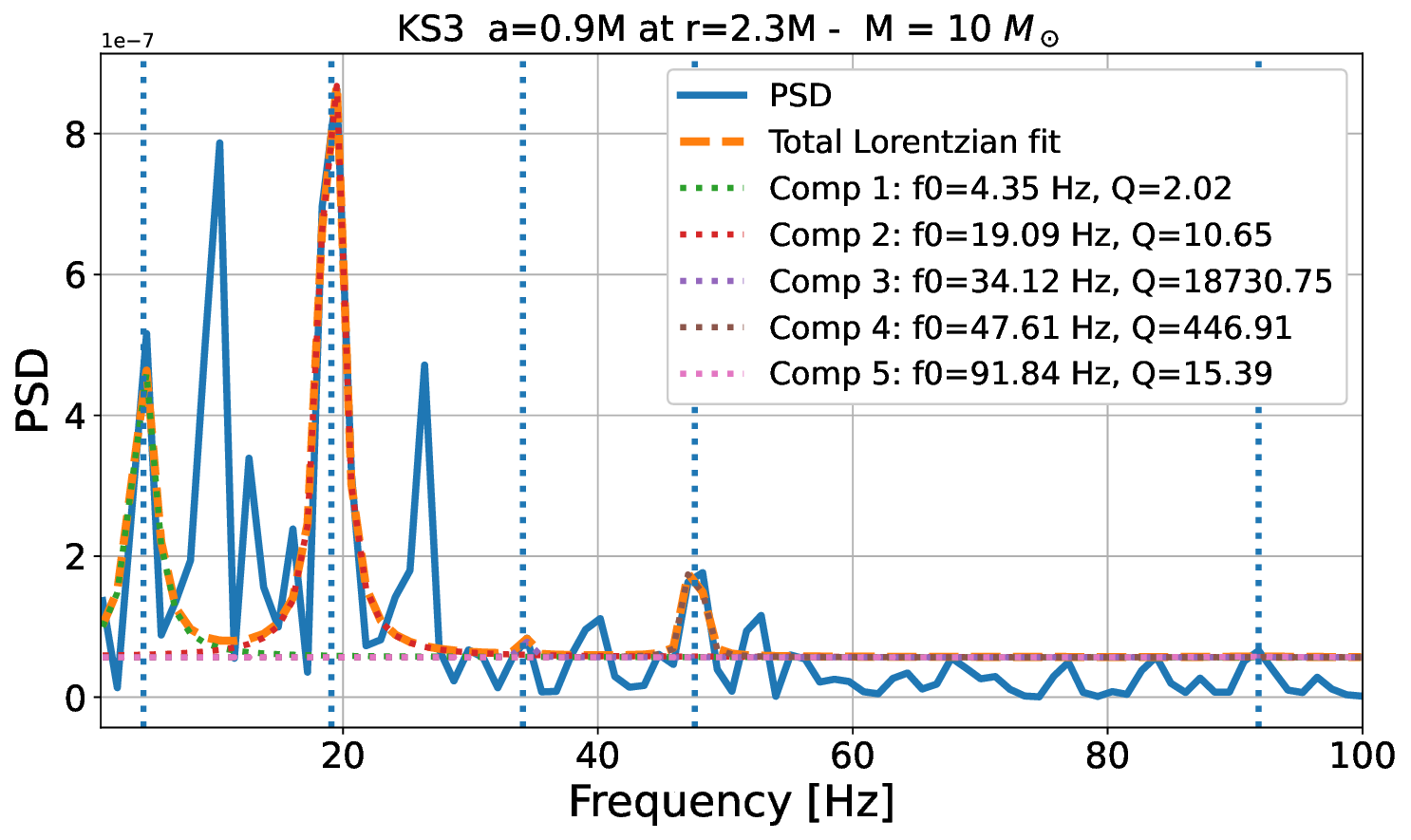}\\
\includegraphics[width=8.cm,height=7.0cm]{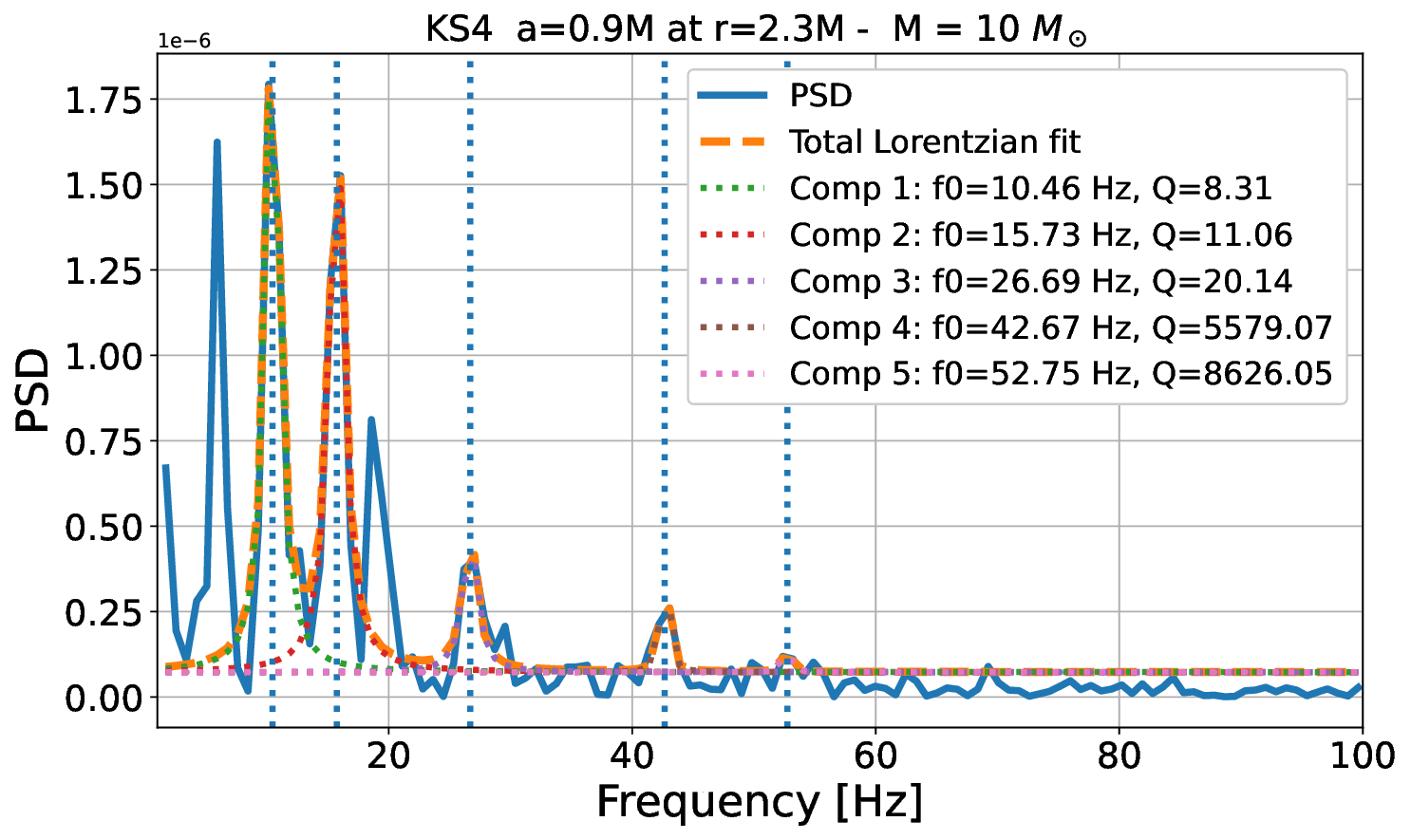} 
\caption{PSD analyses and Lorentzian fits for the rapidly rotating Kerr--Sen and Kerr black holes with $a=0.9M$, including different Kerr--Sen models produced for different values of the deformation parameter, are computed from the mass-accretion rate measured at $r=2.3M$ in the strong gravitational field. The frequencies are scaled by assuming a black hole mass of $10M_{\odot}$ and are shown in Hz. The fitted Lorentzian peaks reveal the dominant QPO-like oscillation modes and show how the Kerr--Sen deformation changes the frequency locations, coherence properties, and harmonic structures relative to the Kerr black hole model.}\label{QPOs_a09}
\end{figure*}

The PSD and Lorentzian fit analyses for the KS1 model are shown in Fig.~\ref{QPOs_a09}. The dominant Lorentzian components occur approximately at $10.15$, $17.48$, $27.11$, $49.40$, and $82.91~\mathrm{Hz}$. The low-frequency components at $10.15$ and $17.48~\mathrm{Hz}$ show that the oscillations of the shock cone excite more than one mode in the inner accretion flow. The component around $27.11~\mathrm{Hz}$ is particularly important because the ratio of these frequencies, $27.11/17.48 \simeq 1.55$, gives an approximate 3:2 ratio. These types of resonance conditions are frequently discussed in the context of black-hole QPO phenomenology \cite{Remillard:2006fc, Abramowicz:2001bi, Varniere:2018zea}. Such a near-commensurable structure may be the result of nonlinear couplings between the modes trapped inside the shock cone. The high-frequency peaks at $49.40$ and $82.91~\mathrm{Hz}$ show that the Kerr--Sen deformation can excite higher harmonics or secondary oscillation modes. The observability of these peaks depends entirely on their strength and coherence. Since the quality factors of the numerically computed peaks in this model are $5.34$ and above, these peaks can be regarded as coherent or strongly coherent. This is one of the required results for observational studies.

For the KS2 model shown again in Fig.~\ref{QPOs_a09}, the dominant Lorentzian components are numerically computed to be approximately around $11.17$, $17.68$, $35.89$, $43.63$, and $62.75~\mathrm{Hz}$. When the frequencies obtained in this model are compared with those of the KS1 model, they produce a clearer harmonic organization. The ratio between $17.68$ and $11.17~\mathrm{Hz}$ is approximately $1.58$, which forms an approximate 3:2 ratio. In addition, the ratio $35.89/17.68 \simeq 2.03$ forms a 2:1 relation, which is also among the observed harmonics. This shows that the moderate Kerr--Sen deformation excites a hierarchy of modes. The peaks formed in this case are thought to arise not only from the modes produced by the fundamental radial and azimuthal oscillations, but also as a result of their nonlinear couplings with each other. The appearance of both the 3:2 and 2:1 ratios in this model makes the KS2 model particularly interesting for revealing QPO-like temporal signatures and for proposing a physical mechanism for the observed QPOs. This is because such frequency relations are used as indicators of resonant behavior in black hole accretion systems.

For the KS3 model shown in Fig.~\ref{QPOs_a09}, the PSD and Lorentzian analyses show that the peaks occur around $4.35$, $19.09$, $34.12$, $47.61$, and $91.84~\mathrm{Hz}$. When compared with the KS1 and KS2 models, the frequency distribution is more irregular, showing that the stronger deformation changes the oscillatory response of the accretion flow. The low-frequency component at $4.35~\mathrm{Hz}$ may be a result of the global modulation of the shock cone. On the other hand, the components at $19.09$, $34.12$, and $47.61~\mathrm{Hz}$ may have formed as a result of the excitation of the modes trapped inside the shock cone in the post-shock region. An approximately 2:1 ratio is observed between $91.84$ and $47.61~\mathrm{Hz}$, since $91.84/47.61 \simeq 1.93$. In general, when the frequency distributions are considered, the frequency distribution is less harmonic than that of KS2. This indicates that stronger deformation may lead to more complex and less regular nonlinear couplings.

In the KS4 model shown in Fig.~\ref{QPOs_a09}, the fitted peaks are numerically observed to occur approximately at $10.46$, $15.73$, $26.69$, $42.67$, and $52.75~\mathrm{Hz}$. This model appears as one of the models that produces the clearest near-resonant structure among the models considered so far. For example, the ratio $15.73/10.46 \simeq 1.50$ gives a value very close to the 3:2 relation. The ratio $52.75/26.69 \simeq 1.98$ produces a resonance state very close to 2:1. Thus, the KS4 model, which describes a strongly deformed case, supports the formation of multiple coupled oscillation modes. This behavior is consistent with the strong temporal variations previously observed from the mass-accretion-rate signal.

When the PSD analysis results obtained from all Kerr--Sen models are compared with the Kerr reference case, it is seen that the temporal structure of the accretion flow is clearly modified due to the effects produced by the deformed spacetime. The Kerr reference model produces several characteristic components; however, the Kerr--Sen models shift the locations of these components, change the coherence properties of the oscillation modes, and produce near-harmonic frequency ratios in different combinations. In particular, while the KS1 and KS2 models preserve relatively organized low-frequency structures, including the 3:2 ratio, the KS3 model produces a more irregular distribution of peaks in the case where the deformation is more effective. In the KS4 model, where the deformation parameter is strongest for the rapidly rotating black hole case, clear near-resonant behaviors appear again, and as a result, the 3:2 and 2:1 harmonic states are formed. These differences between the models show that the Kerr--Sen deformation does not only change the hydrodynamical structure of the shock cone, but also modifies the characteristic oscillations calculated from the mass-accretion rate. Thus, the PSD and Lorentzian analyses provide a direct timing diagnostic for distinguishing Kerr--Sen accretion flows from the Kerr reference solution.

\begin{figure*}[htbp]
\centering
\includegraphics[width=8.cm,height=7.0cm]{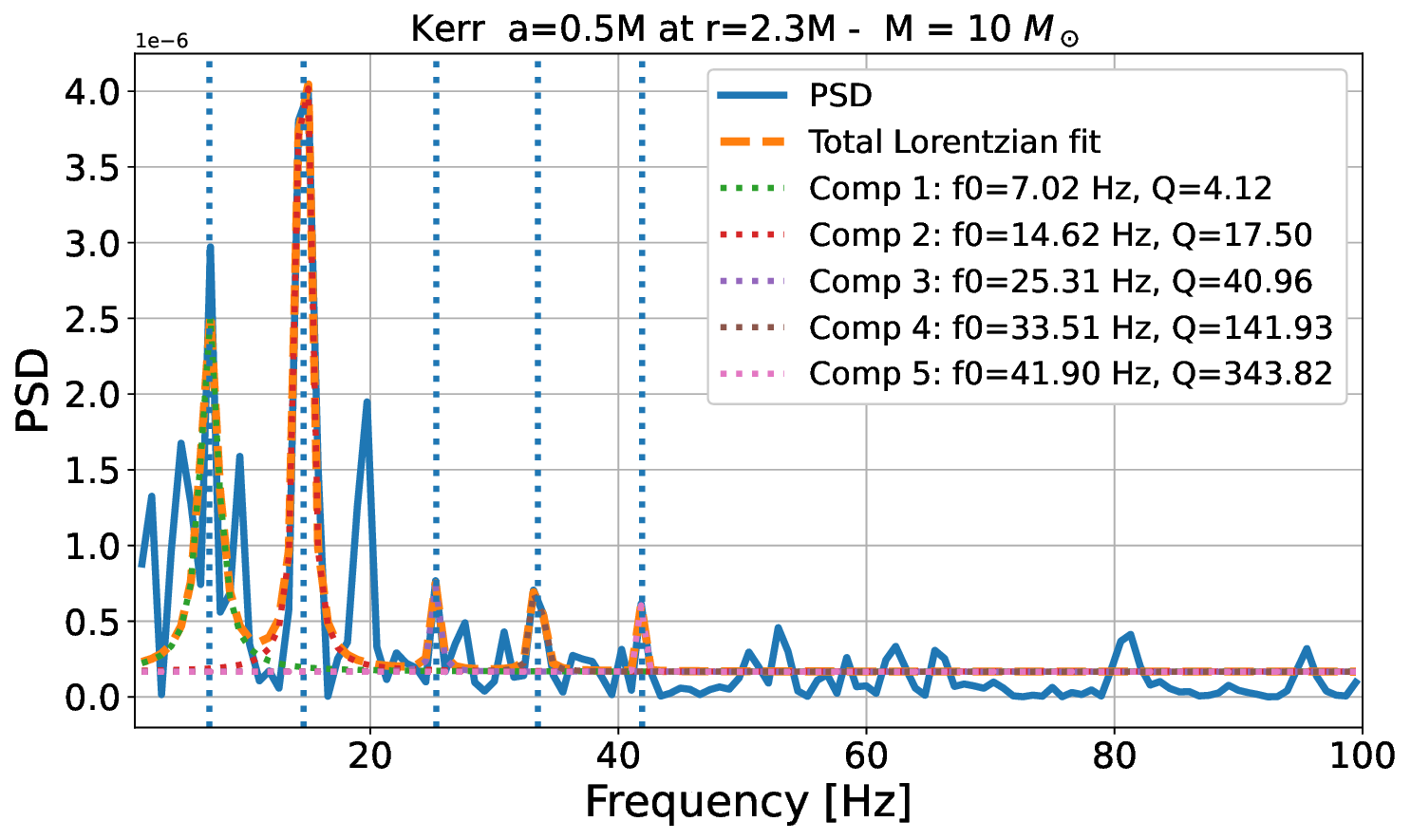}
\includegraphics[width=8.cm,height=7.0cm]{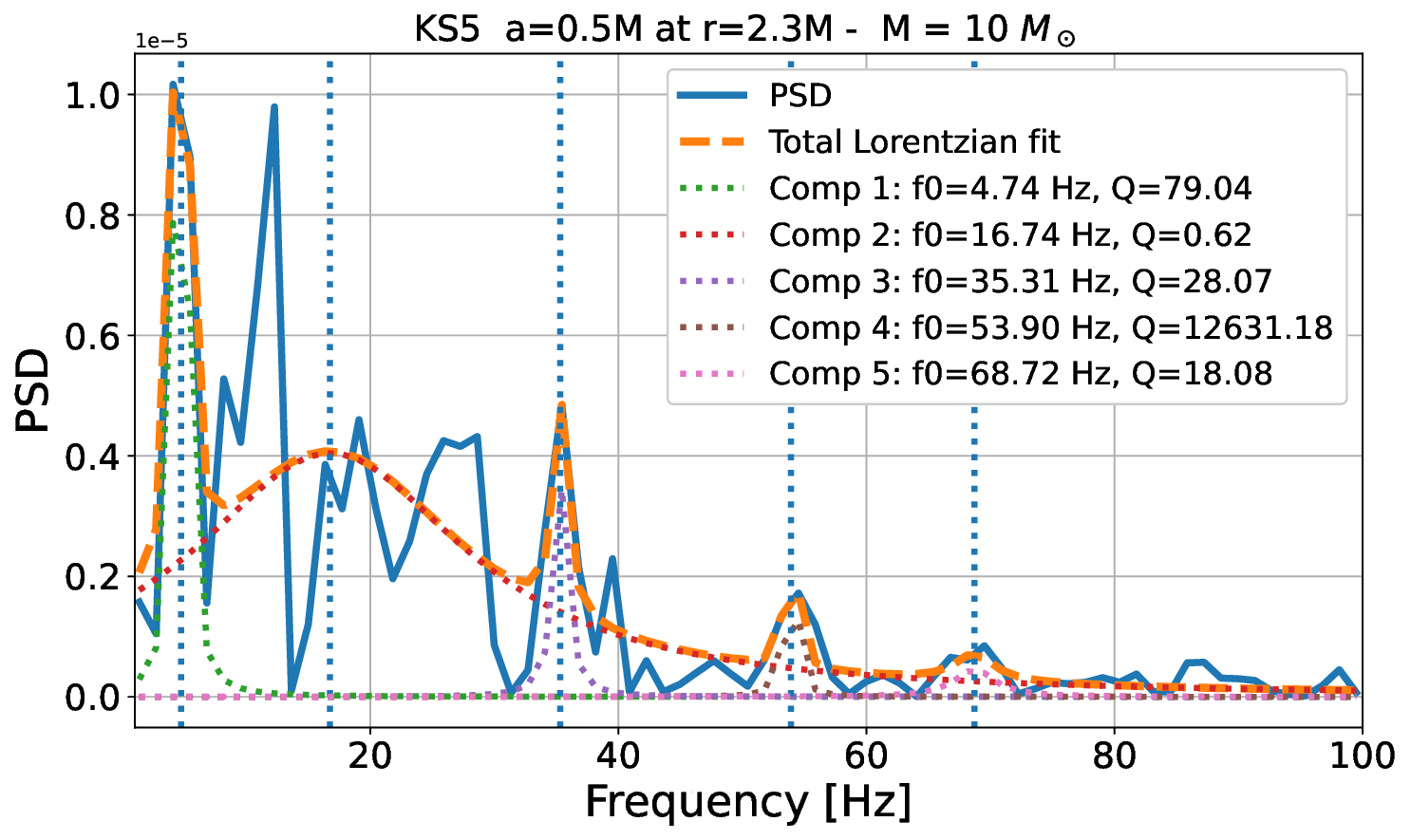} \\
\includegraphics[width=8.cm,height=7.0cm]{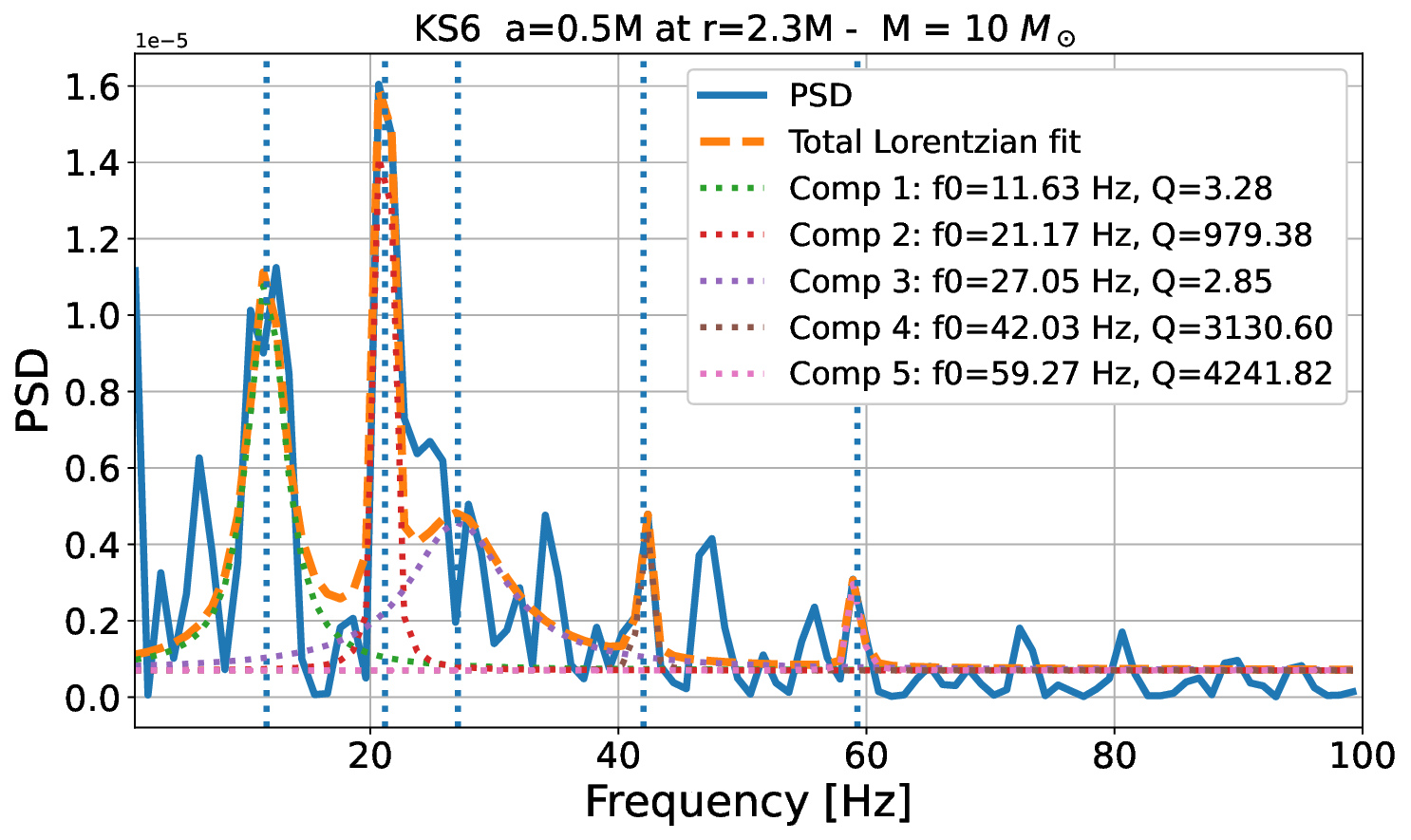}
\caption{It is same as Fig.~\ref{QPOs_a09} but is for Kerr and Kerr–Sen models with the black hole spin $a = 0.5M$. The fitted peaks reveal the dominant QPO-like modes and their deformation-dependent harmonic structure.}\label{QPOs_a05}
\end{figure*}

Fig.~\ref{QPOs_a05} shows the PSD analyses and multi-component Lorentzian fits computed from the mass-accretion rate for the Kerr and Kerr--Sen models in the case where the black hole has a moderate spin parameter, $a=0.5M$. When compared with the rapidly rotating black hole models given in Fig.~\ref{QPOs_a09}, the moderate-spin models produce different temporal structures. This suggests that the interaction between the Kerr--Sen deformation parameter and the black hole spin significantly affects the excitation, coherence properties, and distribution of the oscillation modes. Since the black hole rotates more slowly in the models shown in Fig.~\ref{QPOs_a05}, the effect of the spacetime deformation becomes more dominant in the PSD analyses. This causes stronger shifts in the locations of the peaks and changes the harmonic content of the mass-accretion-rate variability.

In the KS5 model shown in Fig.~\ref{QPOs_a05}, the dominant Lorentzian components are seen to occur around $4.74$, $16.74$, $35.31$, $53.90$, and $68.72~\mathrm{Hz}$. The low-frequency component at $4.74~\mathrm{Hz}$ has a very large quality factor, $Q=79.04$. This implies that this peak is highly coherent. This peak may have emerged as a result of the slow variation of the global modulation of the shock structure. The component at $16.74~\mathrm{Hz}$ has a very small quality factor. This shows that this peak is very broad and has very low coherence. This means that the observability of this peak may be difficult. The high-frequency peaks at $35.31$, $53.90$, and $68.72~\mathrm{Hz}$ show that the Kerr--Sen deformation excites additional peaks in a wide frequency band. In particular, the ratio $53.90/35.31 \simeq 1.53$ produces an approximate 3:2 harmonic. This behavior demonstrates that possible nonlinear couplings may exist between the modes trapped inside the shock cone in this model. On the other hand, the ratio $68.72/35.31 \simeq 1.95$ forms an almost 2:1 harmonic state. Thus, the KS5 model shows that both low-frequency coherent behavior and near-resonant harmonic structures may be possible.

For the KS6 model shown in Fig.~\ref{QPOs_a05}, the Lorentzian peak components are observed to occur at $11.63$, $21.17$, $27.05$, $42.03$, and $59.27~\mathrm{Hz}$. This model represents the case with the strongest Kerr--Sen deformation for the moderately rotating black hole model with $a=0.5M$. The PSD structure formed in this case shows that more complex oscillation behavior occurs. The low-frequency component at $11.63~\mathrm{Hz}$ has a moderate quality factor. In contrast, the peak at $21.17~\mathrm{Hz}$ produces a very large quality factor, indicating a strongly coherent behavior. The ratio $21.17/11.63 \simeq 1.82$ does not produce any simple 3:2 or 2:1 relation, but it shows that the deformation changes the frequency spacing between the excited modes. The component at $42.03~\mathrm{Hz}$ produces the ratio $42.03/21.17 \simeq 1.99$, which is very close to a 2:1 harmonic structure. In addition, $59.27~\mathrm{Hz}$ gives $59.27/21.17 \simeq 2.80$, indicating that higher-order harmonics or nonlinear couplings may exist. Thus, the KS6 model shows that, in the strongly deformed case, a mixture of coherent peaks can form, together with broader oscillation components and near-harmonic structures.

When the Kerr--Sen models in Fig.~\ref{QPOs_a05} are compared with the Kerr model, the effect of the deformation parameter is clearly seen in the temporal behavior. While the Kerr reference model produces several characteristic peaks, in the KS5 and KS6 models the frequency locations are shifted, the coherence properties of the fitted components are changed, and new near-harmonic combinations are formed. In the KS5 model, the PSD produces a very coherent low-frequency peak together with the 3:2 and 2:1 harmonic states at higher frequencies. In the KS6 model, the strong deformation causes the formation of a more complex structure. At the same time, a clear 2:1 harmonic state is also produced. The differences between the Kerr and Kerr--Sen models show that the Kerr--Sen deformation not only modifies the hydrodynamical shock structure and changes the mass-accretion rate, but also leads to the formation of QPO-like characteristic behavior. Thus, Fig.~\ref{QPOs_a05} shows that the moderately rotating Kerr--Sen models provide a strong timing diagnostic for distinguishing deformed Kerr--Sen accretion flows from the Kerr reference solution.

\section{Mass-Scaled Observational Signatures of Kerr–Sen QPO-Like Oscillations}
\label{compare1}
In this section, we establish a connection between the QPO-like oscillation modes numerically computed in the Kerr--Sen accretion simulations and observed black-hole QPOs over different mass ranges through a mass-scaled comparison. The characteristic frequencies extracted from the PSD and Lorentzian analyses are computed for the reference black-hole mass used throughout the paper, $M=10M_{\odot}$. From this point on, these computed QPOs are rescaled by using the observed black-hole masses, allowing a direct comparison between the numerically computed QPO-like behaviors and the observational results. The rescaled numerical frequency for an observed black-hole mass is calculated as 

\begin{equation} f_{\rm num}(M_{\rm obs}) = f_{\rm num}(10M_{\odot}) \times \frac{10M_{\odot}}{M_{\rm obs}} . 
\label{mass_sca}
\end{equation} 

\noindent
Thus, by means of mass scaling, we recalculate the numerically obtained frequencies and obtain the opportunity to directly compare them with stellar-mass, intermediate-mass, and supermassive black-hole candidates. By applying the same scaling relation to different observational sources, we aim to identify which Kerr--Sen models produce frequency ranges and harmonic ratios compatible with the observed timing signatures.

\subsection{Comparison with Stellar-Mass Black-Hole QPOs}
\label{compare2}
First, we compare the observed stellar-mass black-hole sources with the QPO-like behaviors numerically computed in the Kerr--Sen models, and we show in Table~\ref{tab:stellar_com}  which model may be compatible with which observational results. Among the stellar-mass black-hole systems considered here, the source GRS~1915+105 shows the closest relation with the Kerr--Sen models. The observed frequencies in this source occur around $41$ and $67~\mathrm{Hz}$. The calculated mass of this black hole is approximately $12.4M_{\odot}$. In the KS1 model, when the black-hole mass is taken as $10M_{\odot}$, the coherent frequencies appear around $49.40$ and $82.91~\mathrm{Hz}$. If these numerically computed frequencies are rescaled for $M=12.4M_{\odot}$, the resulting QPOs are found to occur at $39.84$ and $66.86~\mathrm{Hz}$. These values are very close to the observed $41$ and $67~\mathrm{Hz}$ QPOs. Thus, the KS1 model appears as a model that can produce both the low- and high-frequency components of the source GRS~1915+105.

As seen in Table~\ref{tab:stellar_com}, the second closest agreement with the source GRS~1915+105 appears in the results obtained from the KS4 model. The numerical frequency $52.75~\mathrm{Hz}$ is rescaled to $42.54~\mathrm{Hz}$ for a black-hole mass of $12.4M_{\odot}$, which is close to the observed $41~\mathrm{Hz}$ component. This indicates that the KS4 model, which corresponds to the strongest deformation case when the black-hole spin parameter is $a=0.9M$, also appears as a model that can explain one of the observed QPO frequencies of this source. Since both KS1 and KS4 represent the models in which the black hole rotates rapidly, namely the case $a=0.9M$, this comparison provides information about the possible spin dependence of the observed QPOs. The best simultaneous match for GRS~1915+105 is obtained from KS1, which favors the rapidly rotating Kerr--Sen family. This is consistent with the observational expectation that GRS~1915+105 is generally interpreted as a high-spin black-hole system, with an estimated spin of about $0.7$ or higher. Thus, the fact that the numerical models with $a=0.9M$ can explain this source appears as an agreement between the numerical results and the observational results.

The second black-hole source compared in Table~\ref{tab:stellar_com} is IGR~J17091--3624. The QPO observed from this source is around $66~\mathrm{Hz}$. This frequency is consistent with the observed high-frequency peaks in the KS1 model. Since the mass calculated from the observational results for this source varies within a certain range, the numerical frequencies rescaled using these masses vary in the range $60.52$--$70.26~\mathrm{Hz}$. Since this range also includes the observed $66~\mathrm{Hz}$ QPO, we can say that the numerical results and the observational results are compatible. Thus, the KS1 model can reproduce the observed QPO of IGR~J17091--3624 within the allowed mass range.

In addition to the rapidly rotating black-hole model KS1, the moderately rotating black-hole model KS5 also provides important information for the source IGR~J17091--3624. In this case, the numerically computed frequency is $68.72~\mathrm{Hz}$. When this frequency is rescaled using the possible observed mass range of this source, the numerical frequency is found to occur around $50.16$--$58.24~\mathrm{Hz}$. These values are lower than the observed $66~\mathrm{Hz}$ QPOs. Therefore, the agreement between the numerical and observational results seen in the KS1 model is not fully observed in the KS5 model. However, the spin value used in the KS5 model, $a=0.5M$, is close to the approximate spin value calculated for this source, $a=0.54M$. Therefore, although KS5 is less successful in matching the observed frequency, it is more consistent with the estimated spin of this source.

\begin{table*}[htbp]
\centering
\caption{Comparison between observed stellar-mass black-hole QPOs and the mass-scaled numerical Kerr--Sen frequencies. The numerical frequencies are originally computed for $M=10M_{\odot}$ and are rescaled using Eq.~\ref{mass_sca}.} 
\label{tab:stellar_com} 
\begin{tabular}{lcccccc}
\hline Source & Observed mass & Observationally & Observed QPOs & Numerical model & Numerical frequencies & Scaled numerical \\
& & estimated spin & [Hz] & & at $10M_{\odot}$ [Hz] & frequencies [Hz] \\ \hline 
GRS~1915+105 & $12.4M_{\odot}$ \cite{Sreehari:2020jge} & about $0.7$ or higher & $41,\;67$ \cite{Sreehari:2020jge, Zhang:2022epl} & KS1 $(a=0.9M)$ & $49.40,\;82.91$ & $39.84,\;66.86$ \\ 
GRS~1915+105 & $12.4M_{\odot}$ & about $0.7$ or higher & $41$ \cite{Belloni:2013qka} & KS4 $(a=0.9M)$ & $52.75$ & $42.54$ \\  IGR~J17091--3624 & $11.8$--$13.7M_{\odot}$ \cite{Iyer:2015tga} & $\sim 0.54$ \cite{Debnath:2025sep} & $66$ \cite{2012AA540L4R} & KS1 $(a=0.9M)$ & $82.91$ & $60.52$--$70.26$ \\ IGR~J17091--3624 & $11.8$--$13.7M_{\odot}$ \cite{Iyer:2015tga} &  $\sim 0.54$ \cite{Debnath:2025sep} & $66$ \cite{2012AA540L4R} & KS5 $(a=0.5M)$ & $68.72$ & $50.16$--$58.24$ \\ 
\hline
\end{tabular}
\end{table*}

As a result, it is seen from Table~\ref{tab:stellar_com} that the KS1 model shows the strongest agreement with the observed stellar-mass black-hole sources given in Table~\ref{tab:stellar_com}. It explains the $41$ and $67~\mathrm{Hz}$ QPO pair of GRS~1915+105 after mass scaling and also reproduces the $66~\mathrm{Hz}$ QPO of IGR~J17091--3624 within the allowed mass interval. While the KS4 model can explain the $41~\mathrm{Hz}$ component observed from the source GRS~1915+105, the KS5 model appears as a more suitable model for the source IGR~J17091--3624. Although the numerical frequencies obtained from the KS5 model are not in one-to-one agreement with the QPO frequencies observed from this source, the resulting numerical frequency range is of the type that can explain the QPOs observed from this source. Therefore, we predict that the approximately measured spin of the source IGR~J17091--3624 being at a moderate level can also be numerically confirmed. Thus, the observational sources given in Table~\ref{tab:stellar_com} appear as observations that can be broadly compatible with the Kerr–Sen frequency range considered here.

\subsection{Comparison with Intermediate-Mass Black-Hole QPO Candidates}
\label{compare3}
In this section, we compare the observed properties of intermediate-mass black-hole candidates with the numerical results obtained from Kerr--Sen gravity. These systems are particularly important because, as a result of inverse mass scaling, the QPO frequencies observed from these black-hole sources are lower than those observed from stellar-mass black holes. Thus, if the QPO-like modes obtained from the Kerr--Sen simulations are physically relevant, the same numerical frequencies computed for $M=10M_{\odot}$ shift to the Hz or mHz level. In Table~\ref{tab:intermediate_com}, we present the comparison between the numerical and observational results for the well-known intermediate-mass black-hole candidates M82~X--1 and NGC~5408~X--1.

The source M82~X--1 given in Table~\ref{tab:intermediate_com} is one of the important intermediate-mass black-hole candidates because twin QPOs at $3.32$ and $5.07~\mathrm{Hz}$ have been observed from this source. These two frequencies form an approximate 3:2 harmonic ratio. Therefore, this source is particularly important for testing not only whether the Kerr--Sen model can produce the QPO frequencies, but also whether the numerically computed QPO frequencies can form a near-resonant timing structure. As seen in Table~\ref{tab:intermediate_com}, the observed mass interval of M82~X--1 is $140$--$660M_{\odot}$. When the frequencies $49.40$ and $82.91~\mathrm{Hz}$ obtained in the KS1 model are rescaled according to the observed mass range of this source, the frequencies are found to be in the ranges $0.75$--$3.53~\mathrm{Hz}$ and $1.26$--$5.92~\mathrm{Hz}$. The numerical QPO-like peaks calculated for this source are compatible with the observed $3.32$ and $5.07~\mathrm{Hz}$ QPOs. Thus, the KS1 model can reproduce the observed QPO pair of M82~X--1 within the allowed mass interval.

The KS5 model with the moderate spin parameter $a=0.5M$ also provides important comparison results for the source M82~X—1 as seen in Table~\ref{tab:intermediate_com}. In the KS5 model, the frequencies obtained for a black hole with $M=10M_{\odot}$ are $53.90$ and $68.72~\mathrm{Hz}$. When these frequencies are rescaled according to the observed mass range of M82~X--1, the QPO frequencies become $0.82$--$3.85~\mathrm{Hz}$ and $1.04$--$4.91~\mathrm{Hz}$. These values fall within the range of the observed $3.32$ and $5.07~\mathrm{Hz}$ QPOs. Although the upper numerical frequency is slightly different from the observed $5.07~\mathrm{Hz}$ value, the fact that the rescaled numerical frequencies lie in the same observed band reveals a strong agreement between the numerical and observational results. Thus, the KS1 and KS5 models can explain the observed temporal behavior of the source M82~X--1. However, the interpretation is different for the two models. KS1 provides a strong frequency match from the rapidly rotating Kerr--Sen family, whereas KS5 provides a comparison that is more consistent with the observationally estimated spin interval of about $0.05$--$0.6$. Thus, for the source M82~X--1, the KS5 model may be considered physically more suitable because of the agreement between the numerical and observational spin values. On the other hand, since the QPO-like peaks obtained from the KS1 model show excellent agreement with the observations, this model also appears as an important model that can be used to explain this observed source.

As seen in Table~\ref{tab:intermediate_com}, the second intermediate-mass black-hole candidate compared is NGC~5408~X--1. The QPO frequencies observed from this source are in the range $0.010$--$0.020~\mathrm{Hz}$. The corresponding observed black-hole masses vary in the range $1000$--$9000M_{\odot}$. When the numerically computed frequencies $10.15$ and $17.48~\mathrm{Hz}$ of the rapidly rotating black-hole model KS1 are rescaled using the observed masses, the numerical results for the source NGC~5408~X--1 are found to be in the ranges $0.011$--$0.102~\mathrm{Hz}$ and $0.019$--$0.175~\mathrm{Hz}$. These intervals include the observed $0.010$--$0.020~\mathrm{Hz}$ QPO range. In particular, the agreement between the numerical and observational results is stronger toward the higher end of the observed black-hole mass range. This suggests that the rapidly rotating KS1 model can naturally explain the temporal behavior of the source NGC~5408~X--1 if the source has a mass closer to the upper part of the estimated intermediate-mass range.

The KS5 model, whose details are given in Table~\ref{tab:intermediate_com}, produces rescaled frequencies at the mHz level for the source NGC~5408~X--1. When the numerical frequencies $4.74$ and $16.74~\mathrm{Hz}$, computed for a black-hole mass of $M=10M_{\odot}$, are rescaled using the observed black-hole masses in the range $1000$--$9000M_{\odot}$, the frequencies are found to occur in the ranges $0.005$--$0.047~\mathrm{Hz}$ and $0.019$--$0.167~\mathrm{Hz}$. The second scaled frequency interval includes the observed $0.010$--$0.020~\mathrm{Hz}$ range, while the first interval extends below the observed band. Thus, the KS5 model can explain only part of the observed temporal behavior of this source. However, the KS5 model does not naturally reproduce the observed range completely, as in the KS1 model. This implies that the KS1 model may be a stronger model for explaining the source NGC~5408~X--1.

As reported in Table~\ref{tab:intermediate_com}, the spin of the source NGC~5408~X--1 is not observationally known. Therefore, by using the numerical--observational comparison performed here, a theoretical prediction can be made for the spin of this source. If the observed $0.010$--$0.020~\mathrm{Hz}$ QPOs of NGC~5408~X--1 are considered within the Kerr--Sen framework, and if the mass of this source is assumed to be close to the upper limit of the observed $1000$--$9000M_{\odot}$ range, then the KS1 model emerges as the best model that can explain this source. This suggests that NGC~5408~X--1 may be a rapidly rotating black hole with a spin parameter around $a=0.9M$. In other words, although the KS5 model with spin parameter $a=0.5M$ still produces QPO-like frequencies in the mHz range, the agreement of this model with the observations is weaker and only partial. Thus, the timing comparisons show that NGC~5408~X--1 is broadly compatible with the rapidly rotating Kerr--Sen model and therefore suggest that this black hole may be a rapidly rotating black hole.

\begin{table*}[htbp] 
\centering 
\caption{Comparison between observed intermediate-mass black-hole QPO candidates and the mass-scaled numerical Kerr--Sen frequencies. The numerical frequencies are originally computed for $M=10M_{\odot}$ and are rescaled using Eq.~\ref{mass_sca}.} 
\label{tab:intermediate_com} 
\begin{tabular}{lcccccc} 
\hline 
Source & Observed mass & Observationally & Observed QPOs & Numerical model & Numerical frequencies & Scaled numerical \\ 
& & estimated spin & [Hz] & & at $10M_{\odot}$ [Hz] & frequencies [Hz] \\ \hline
M82~X--1 & $140$--$660M_{\odot}$ \cite{Mucciarelli:2005jx}& about $0.05$--$0.6$ & $3.32,\;5.07$ \cite{Pasham:2014ybe}& KS1 $(a=0.9M)$ & $49.40,\;82.91$ & $0.75$--$3.53,\;1.26$--$5.92$ \\ 
M82~X--1 & $140$--$660M_{\odot}$ \cite{Mondal:2022hle}& about $0.05$--$0.6$ & $3.32,\;5.07$ & KS5 $(a=0.5M)$ & $53.90,\;68.72$ & $0.82$--$3.85,\;1.04$--$4.91$ \\ 
NGC~5408~X--1 & $1000$--$9000M_{\odot}$ \cite{2012RAA....12.1597H}& uncertain & $0.010$--$0.020$ \cite{Pasham:2012thw}& KS1 $(a=0.9M)$ & $10.15,\;17.48$ & $0.011$--$0.102,\;0.019$--$0.175$ \\ NGC~5408~X--1 & $1000$--$9000M_{\odot}$ & uncertain & $0.010$--$0.020$ & KS5 $(a=0.5M)$ & $4.74,\;16.74$ & $0.005$--$0.047,\;0.019$--$0.167$ \\ \hline 
\end{tabular} 
\end{table*}

\subsection{Comparison with Supermassive Black-Hole QPOs}
\label{compare4}
Finally, we compare the numerically computed Kerr--Sen QPO-like frequencies with the observed supermassive black-hole systems reported in the literature. These comparisons and their results are given in Table~\ref{tab:supermassive_com}. For this purpose, we first calculate the QPO-like frequencies in the Kerr--Sen models, which are computed for black holes with mass $M=10M_{\odot}$, for the supermassive black-hole sources by using mass scaling and the observed masses of the sources. Thus, if the QPO-like frequencies found in the Kerr--Sen models are physically relevant, the same numerical modes also appear in supermassive black-hole sources in the range $10^{-5}$--$10^{-4}~\mathrm{Hz}$.

As seen in Table~\ref{tab:supermassive_com}, we first examine the supermassive black-hole source RE~J1034+396. The observed QPO frequencies of this source fall approximately in the range $(2.47$--$2.83)\times10^{-4}~\mathrm{Hz}$, while the observed black-hole mass is reported to be in the range $(1$--$4)\times10^{6}M_{\odot}$. When the Kerr--Sen model KS1, namely the model with spin parameter $a=0.9M$, is considered, the numerically computed frequencies for a black-hole mass of $M=10M_{\odot}$ are $49.40$ and $82.91~\mathrm{Hz}$. If these frequencies are rescaled using the observed mass of the source RE~J1034+396, the frequencies are found to vary in the ranges $(1.24$--$4.94)\times10^{-4}~\mathrm{Hz}$ and $(2.07$--$8.29)\times10^{-4}~\mathrm{Hz}$. These intervals include the observed QPO range of the source RE~J1034+396. Thus, the rapidly rotating KS1 model can reproduce the observed QPO frequency of this source within the allowed mass interval.

At the same time, as seen in Table~\ref{tab:supermassive_com}, the numerical QPO-like behaviors obtained in the moderately rotating black-hole model KS5 are also capable of explaining the results obtained from the source RE~J1034+396. If the numerical QPOs obtained in the KS5 model, $53.90$ and $68.72~\mathrm{Hz}$, are rescaled, it is numerically computed that they can exhibit QPO-like behaviors in the ranges $(1.35$--$5.39)\times10^{-4}~\mathrm{Hz}$ and $(1.72$--$6.87)\times10^{-4}~\mathrm{Hz}$. These rescaled frequency intervals are compatible with the observed QPO range. Thus, both the KS1 and KS5 models are capable of explaining the observed temporal behavior of the source RE~J1034+396. On the other hand, the spin of this source has not been fully calculated. Therefore, due to the agreement between the numerically computed QPOs and the observational results, an estimate can be made for the spin parameter of this black-hole source. In other words, when the agreement of the numerically modeled Kerr--Sen models KS1 and KS5 with the observational results is taken into account, it is predicted that the spin parameter of this source may lie in the range from moderate to rapid rotation.

The second source whose properties are given in Table~\ref{tab:supermassive_com} is 1H~0707--495. The QPO frequency observed from this source is $2.63\times10^{-4}~\mathrm{Hz}$, and the observed black-hole mass range is $(2.0$--$5.2)\times10^{6}M_{\odot}$. If we consider the KS1 model, the numerically computed frequency $82.91~\mathrm{Hz}$ is rescaled to $(1.59$--$4.15)\times10^{-4}~\mathrm{Hz}$. This appears as the numerical QPO-like behavior predicted for the source 1H~0707--495 depending on the observed mass of the source. This numerically computed interval also contains the observed QPO frequency. Thus, when the observed mass range of the source 1H~0707--495 is taken into account, the KS1 model appears as a model that can explain the observed temporal properties of this source.

Again, the comparison of the observed results from the same source, 1H~0707--495, with the KS4 model is given in Table~\ref{tab:supermassive_com}. In this model, when the numerically obtained frequency $52.75~\mathrm{Hz}$ is rescaled using the observed mass of the source, QPO-like behaviors are found to occur in the range $(1.01$--$2.64)\times10^{-4}~\mathrm{Hz}$. This range reaches the observed value $2.63\times10^{-4}~\mathrm{Hz}$ at the upper edge of the scaled interval. Thus, the KS4 model can also explain the observed QPO frequency of 1H~0707--495. Since both the KS1 and KS4 models are results of the rapidly rotating case $(a=0.9M)$ of the Kerr--Sen family, the agreement between the models for this source shows that this source has a rapidly rotating black hole, as also reported in the observations. Thus, the source 1H~0707--495 appears as one of the strongest supermassive black-hole sources in which the observed QPO can be interpreted in terms of rapidly rotating Kerr--Sen models.
 
The third supermassive black-hole source given in Table~\ref{tab:supermassive_com} is ESO~113--G010. The two QPO frequencies observed from this source are $6.79\times10^{-5}~\mathrm{Hz}$ and $1.24\times10^{-4}~\mathrm{Hz}$. The black-hole mass calculated from the observational results is approximately $7\times10^{6}M_{\odot}$. If the frequencies $53.90$ and $68.72~\mathrm{Hz}$ obtained in the KS5 model are recalculated according to the observed mass of this black hole, the numerically obtained frequencies become $7.70\times10^{-5}~\mathrm{Hz}$ and $9.82\times10^{-5}~\mathrm{Hz}$. These numerical results also fall within the observed QPO band. In particular, while the first scaled frequency is very close to the observed $6.79\times10^{-5}~\mathrm{Hz}$ component, the second scaled frequency fits well within the observed band. This behavior demonstrates that the KS5 model, which has a moderate spin parameter $a=0.5M$, appears as a model that can naturally explain the temporal behavior of the source ESO~113--G010.

Again, the numerical results obtained from the KS4 model for the same source, ESO~113--G010, given in Table~\ref{tab:supermassive_com}, are also comparable. When the numerically computed frequencies $42.67$ and $52.75~\mathrm{Hz}$ in the KS4 model are rescaled, the numerical QPOs predicted for the supermassive black-hole source ESO~113--G010 occur in the range $6.10\times10^{-5}~\mathrm{Hz}$ and $7.54\times10^{-5}~\mathrm{Hz}$. These numerical results are very close to the lower limit of the observed QPO frequencies. However, they do not reproduce the higher observed frequency as well as the KS5 model. Thus, although the KS4 model is a model that can partially explain the observed temporal behavior of ESO~113--G010, the KS5 model shows stronger agreement with the observational results. Again, for the source ESO~113--G010, there is no observationally calculated or estimated spin parameter. Therefore, based on the agreement that we have numerically obtained here, it is predicted that the spin of ESO~113--G010 may be closer to $a=0.5M$. In other words, this black hole may be a moderately rotating black hole.

\begin{table*}[htbp] 
\centering 
\scriptsize 
\setlength{\tabcolsep}{4pt} 
\renewcommand{\arraystretch}{1.05}
\caption{Comparison between observed supermassive black-hole QPOs and the mass-scaled numerical Kerr--Sen frequencies. The numerical frequencies are originally computed for $M=10M_{\odot}$ and are rescaled using Eq.~\ref{mass_sca}.}
\label{tab:supermassive_com} 
\begin{tabular}{lcccccc} 
\hline 
Source & Observed mass & Observationally & Observed QPOs & Numerical model & Numerical frequencies & Scaled numerical \\ 
& & estimated spin & [Hz] & & at $10M_{\odot}$ [Hz] & frequencies [Hz] \\ \hline
RE~J1034+396 & $(1$--$4)\times 10^{6}M_{\odot}$ \cite{2010MNRAS.401..507B}& uncertain & $(2.47$--$2.83)\times 10^{-4}$ \cite{Czerny:2016ajj, Xia:2024yuy} & KS1 $(a=0.9M)$ & $49.40,\;82.91$ & $(1.24$--$4.94)\times 10^{-4},\;(2.07$--$8.29)\times 10^{-4}$ \\
RE~J1034+396 & $(1$--$4)\times 10^{6}M_{\odot}$ & uncertain & $(2.47$--$2.83)\times 10^{-4}$ & KS5 $(a=0.5M)$ & $53.90,\;68.72$ & $(1.35$--$5.39)\times 10^{-4},\;(1.72$--$6.87)\times 10^{-4}$ \\
1H~0707--495 & $(2.0$--$5.2)\times 10^{6}M_{\odot}$ \cite{Pan:2016lmr}& high spin \cite{Fabian2012} & $2.63\times 10^{-4}$ \cite{Pan:2016lmr} & KS1 $(a=0.9M)$ & $82.91$ & $(1.59$--$4.15)\times 10^{-4}$ \\ 
1H~0707--495 & $(2.0$--$5.2)\times 10^{6}M_{\odot}$ & high spin & $2.63\times 10^{-4}$ & KS4 $(a=0.9M)$ & $52.75$ & $(1.01$--$2.64)\times 10^{-4}$ \\
ESO~113--G010 & $\simeq 7\times 10^{6}M_{\odot}$ \cite{Cackett:2012sq}& uncertain & $6.79\times 10^{-5},\;1.24\times 10^{-4}$ \cite{Cackett:2012sq, 2020ChAA..44...32Z} & KS5 $(a=0.5M)$ & $53.90,\;68.72$ & $7.70\times 10^{-5},\;9.82\times 10^{-5}$ \\ 
ESO~113--G010 & $\simeq 7\times 10^{6}M_{\odot}$ & uncertain & $6.79\times 10^{-5},\;1.24\times 10^{-4}$ & KS4 $(a=0.9M)$ & $42.67,\;52.75$ & $6.10\times 10^{-5},\;7.54\times 10^{-5}$ \\ 
\hline
\end{tabular} 
\end{table*}

\section{Conclusions}
\label{conc}
In this study, we reveal the hydrodynamical and temporal signatures of BHL accretion around Kerr and Kerr--Sen black holes, and we show the effect of the spacetime correction by comparing the obtained Kerr--Sen solutions with the Kerr solutions. Our main aim here is to reveal how the charge-related Kerr--Sen deformation parameter, together with the black-hole spin, modifies the shock cone formed around the black hole in the strong gravitational field, the variation of the mass-accretion rate, and the QPO-like oscillation modes. By considering two representative spin families, $a=0.9M$ and $a=0.5M$, we show that Kerr--Sen gravity produces measurable imprints in both the accretion morphology and the characteristic timing behavior in the inner accretion region.

The numerical simulations show that the Kerr--Sen deformation changes the morphology of the shock cone, the density distribution inside the shock cone, and the temporal behavior of the mass-accretion rate. In the rapidly rotating black-hole model with $a=0.9M$, the shock-cone profiles formed around the black hole produce solutions that are relatively close or similar to the Kerr solution. This is because the effect of strong rotation and frame dragging is seen to dominate the flow dynamics. Nevertheless, it is revealed that the Kerr--Sen models still produce changes in the density peaks, in the angular shock structure, and in the variation of the accretion rate. On the other hand, in the case where the black-hole spin is moderate, $a=0.5M$, the effect of the deformation parameter of Kerr--Sen gravity becomes more pronounced. In this case, the density inside the shock cone decreases, the boundaries defining the cone in the angular direction are modified, and the mass-accretion-rate signals vary strongly depending on the deformation. These results show numerically that the Kerr--Sen deviation strongly depends on the coupling between the black-hole spin and the deformation parameter.

The PSD analysis and multi-component Lorentzian fits contain the most direct temporal diagnostics of the Kerr--Sen deformation. In the rapidly rotating black-hole case, the KS1 and KS2 models produce more organized QPO-like structures, while also forming approximate 3:2 and 2:1 harmonic ratios. The KS3 model, on the other hand, shows a more irregular frequency distribution. This reveals that, in the case of stronger deformation, more complex nonlinear couplings are formed. At the same time, clear resonant behaviors also appear in the KS4 model, namely ratios close to 3:2 and 2:1. Among the models in which the black-hole spin is moderate, KS5 produces highly coherent low-frequency components, and the ratios of these components also show behaviors very similar to the 3:2 and 2:1 harmonics. On the other hand, KS6, which is also one of the moderately rotating black-hole models, forms a mixture of coherent peaks and broader components. In this case, only the 2:1 harmonic ratio is seen to occur. Thus, the Lorentzian fits show that the Kerr--Sen deformation does not only cause the characteristic frequencies to shift, but also changes their coherence, harmonic organization, and the nonlinear couplings of the modes trapped inside the shock cone.

The numerical QPOs obtained from the Kerr--Sen models are compared with some observed stellar-mass black-hole sources, and after applying inverse mass scaling, it is seen that the results obtained from the Kerr--Sen models are compatible with the observations. The strongest agreement is observed in the KS1 model. For the observed source GRS~1915+105, the numerical frequencies $49.40$ and $82.91~\mathrm{Hz}$ found in the KS1 model are rescaled using the source mass $12.4M_{\odot}$, and the numerically computed QPO-like behaviors are found to occur at $39.84$ and $66.86~\mathrm{Hz}$. These numerical results are highly compatible with the observed $41$ and $67~\mathrm{Hz}$ QPO pair from this source. The numerical QPOs obtained from the KS4 model also explain the $41~\mathrm{Hz}$ component observed in the same source. On the other hand, the QPOs found from the observed source IGR~J17091--3624 are compatible with the QPO interval $60.52$--$70.26~\mathrm{Hz}$ obtained by rescaling the $82.91~\mathrm{Hz}$ frequency calculated in the KS1 model.   Therefore, the numerically computed QPO interval is compatible with the observed $66~\mathrm{Hz}$ QPO from IGR~J17091--3624. Finally, the KS5 model gives a less accurate frequency match but is more consistent with the estimated moderate spin of this source. Thus, the stellar-mass comparison shows that the rapidly rotating Kerr--Sen models, especially KS1, give the strongest frequency-based agreement with the observed stellar-mass black holes, while the KS5 model may be relevant when spin consistency is considered.

For the comparison with intermediate-mass black-hole observations, the Kerr--Sen QPO-like modes are shifted to the Hz--mHz region by using the mass-scaling relation. When the source M82~X--1 is examined, a QPO pair at $3.32$ and $5.07~\mathrm{Hz}$ has been observed from this source. When the numerical frequencies calculated for the rapidly rotating black-hole model KS1 are recalculated with mass scaling by using the observed mass of this source, the frequency ranges $0.75$--$3.53~\mathrm{Hz}$ and $1.26$--$5.92~\mathrm{Hz}$ are obtained. This is seen to be compatible with the frequency range observed from the source M82~X--1. At the same time, the numerical frequencies obtained from the KS5 model for the moderately rotating black hole are also seen to be compatible with the observational results of the same source. In particular, the agreement of the KS5 model results with the observations is physically interesting because the moderate spin of this model is compatible with the observed spin range of the source M82~X--1. On the other hand, the source NGC~5408~X--1 has an observed QPO range of $0.010$--$0.020~\mathrm{Hz}$. Again, when the numerical results in the KS1 model are rescaled by using the estimated mass of the source NGC~5408~X--1, the numerical results of KS1 are seen to fall within the observed QPO range. At the same time, the numerical results found in KS5 partially explain the observed temporal behavior. However, the agreement with the observational results in this model is weaker than in the KS1 model. Thus, the comparisons made for intermediate-mass black holes show that the observational results of the source M82~X--1 is broadly compatible with the numerical results obtained from both the KS1 and KS5 models, while NGC~5408~X--1 is more naturally described by a rapidly rotating Kerr--Sen configuration, 
suggesting possible consistency with a rapidly rotating Kerr–Sen configuration $a=0.9M$ within the assumptions of the present model.

For the observed supermassive black-hole sources, the numerically computed Kerr--Sen frequencies are shifted to the range $10^{-5}$--$10^{-4}~\mathrm{Hz}$ when the expected masses of supermassive black holes are used. This interval corresponds to the frequency band expected for AGN QPOs. For RE~J1034+396, the numerical frequencies obtained from both KS1 and KS5 fall within the observed QPO range of this source, $2.47$--$2.83\times10^{-4}~\mathrm{Hz}$. This suggests that this source may be compatible with a moderate-to-rapidly rotating Kerr--Sen black hole. For the observed source 1H~0707--495, it is calculated that both the KS1 and KS4 models are compatible with the observed QPO frequency $2.63\times10^{-4}~\mathrm{Hz}$ within the mass range predicted for the source. Since both models belong to the rapidly rotating family with $a=0.9M$, this supports the high-spin interpretation of this source. For the source ESO~113--G010, the KS5 model gives the best agreement with the observed QPO band. In particular, the scaled frequency $7.70\times10^{-5}~\mathrm{Hz}$ from the numerical results of KS5 is very close to the observed frequency component $6.79\times10^{-5}~\mathrm{Hz}$. On the other hand, the KS4 model is compatible only with the lower part of the observed frequencies. Thus, from the comparison with supermassive black holes, it is found that the spin of the source RE~J1034+396 lies in the moderate-to-rapid regime, while the source 1H~0707--495 is explained by a rapidly rotating Kerr--Sen model. ESO~113--G010 is more naturally compatible with a moderately rotating Kerr--Sen black hole.

Finally, it has been discussed in detail that QPO-like modes are excited as a result of the BHL accretion flow around the Kerr--Sen black hole, and that the numerically computed QPOs may explain the observational results obtained from stellar-mass, intermediate-mass, and supermassive black-hole systems. In addition, these numerical results may provide predictions about the black-hole mass and spin that have not been fully determined observationally. Thus, the shock-cone oscillation mechanisms that arise in the simulations provide a unified timing interpretation over a wide range of black-hole masses. The Kerr--Sen deformation changes the accretion morphology, modifies the characteristic oscillation modes, and produces different frequency ratios that are compatible with different observed QPO systems. Future work should extend this analysis to higher-resolution simulations, different asymptotic flow velocities, broader regions of the Kerr--Sen parameter space, and fully three-dimensional accretion flows. Such studies will be able to test the robustness of the QPO-like modes that are formed, allow the spins of sources whose spins have not been observationally measured to be identified within a narrower range, and make it possible to reveal more clearly the behaviors related to testing the Kerr--Sen spacetime.

\section*{Acknowledgments}

All numerical simulations were performed using the Phoenix High Performance Computing facility at the American University of the Middle East (AUM), Kuwait.  

\section*{Data Availability Statement}

The data sets generated and analyzed during the current study were produced using high-performance computing resources. These data are not publicly available due to their large size and computational nature, but are available from the corresponding author upon reasonable request.

\bibliographystyle{unsrt} 
\bibliography{KerrsenRef}
\end{document}